\documentclass[twocolumn]{aastex631}
\usepackage{url}
\usepackage{epsf}
\usepackage {threeparttable}

\usepackage{color}
\shorttitle{The nature of radio galaxies at $z \sim 4$}
\shortauthors{Yamamoto et al.}

\begin{document}

\title{A Wide and Deep Exploration of Radio Galaxies with Subaru HSC (WERGS). X. The Massive and Passive Nature of Radio Galaxies at $z \sim 4$}

\email{yamamoto@cosmos.phys.sci.ehime-u.ac.jp}

\author[0009-0008-3501-7773]{Yuta Yamamoto}
\affiliation{Graduate School of Science and Engineering, Ehime University, 2-5 Bunkyo-cho, Matsuyama, Ehime 790-8577, Japan}

\author{Tohru Nagao}
\affiliation{Research Center for Space and Cosmic Evolution, Ehime University, 2-5 Bunkyo-cho, Matsuyama, Ehime 790-8577, Japan}
\affiliation{Amanogawa Galaxy Astronomy Research Center, Kagoshima University, 1-21-35 Korimoto, Kagoshima 890-0065, Japan}

\author{Takuji Yamashita}
\affiliation{National Astronomical Observatory of Japan, 2-21-1 Osawa, Mitaka, Tokyo 181-8588, Japan}

\author{Hisakazu Uchiyama}
\affiliation{National Astronomical Observatory of Japan, 2-21-1 Osawa, Mitaka, Tokyo 181-8588, Japan}

\author{Mariko Kubo}
\affiliation{Astronomical Institute, Tohoku University, 6-3 Aramaki, Aoba-ku, Sendai, Miyagi 980-8578, Japan}

\author[0000-0002-3531-7863]{Yoshiki Toba}
\affiliation{National Astronomical Observatory of Japan, 2-21-1 Osawa, Mitaka, Tokyo 181-8588, Japan}
\affiliation{Academia Sinica Institute of Astronomy and Astrophysics, 11F of Astronomy-Mathematics Building, AS/NTU, No.1, Section 4, Roosevelt Road, Taipei 10617, Taiwan}
\affiliation{Research Center for Space and Cosmic Evolution, Ehime University, 2-5 Bunkyo-cho, Matsuyama, Ehime 790-8577, Japan}

\author{Yuichi Harikane}
\affiliation{Institute for Cosmic Ray Research, The University of Tokyo, 5-1-5 Kashiwanoha, Kashiwa, Chiba 277-8582, Japan}

\author{Kohei Ichikawa}
\affiliation{Department of Physics, School of Advanced Science and Engineering, Faculty of Science and Engineering, Waseda University, 3-4-1 Okubo, Shinjuku, Tokyo 169-8555, Japan}

\author{Masaru Kajisawa}
\affiliation{Graduate School of Science and Engineering, Ehime University, 2-5 Bunkyo-cho, Matsuyama, Ehime 790-8577, Japan}

\author{Akatoki Noboriguchi}
\affiliation{School of General Education, Shinshu University, 3-1-1 Asahi, Matsumoto, Nagano 390-8621, Japan}

\author{Yoshiaki Ono}
\affiliation{Institute for Cosmic Ray Research, The University of Tokyo, 5-1-5 Kashiwanoha, Kashiwa, Chiba 277-8582, Japan}

\author[0000-0002-3866-9645]{Toshihiro Kawaguchi}
\affiliation{Department of Economics, Management and Information Science, Onomichi City University, Hisayamada 1600-2, Onomichi, Hiroshima 722-8506, Japan}

\begin{abstract}

High-$z$ radio galaxies (HzRGs) are considered important objects for understanding the formation and evolution of massive galaxies in the early universe. However, till date, detailed studies of the stellar population of HzRGs such as the star-formation history have been scarce. Therefore, this study conducted a new survey to establish a less-biased sample of HzRGs and consequently investigate their properties. We utilized a sample of $g$-dropout Lyman break galaxies (LBGs) obtained from an optical wide and deep imaging survey made by Subaru Hyper Suprime-Cam (HSC). Based on the cross-matching of this LBG sample with the VLA FIRST radio survey data, we constructed a photometric sample of high-redshift radio galaxies (HzRGs) at  $z \sim 4$ for $\sim$560 deg$^2$ survey field. Consequently, we identified 146 HzRG candidates. To analyze the characteristics of these candidates, we focus on objects exhibiting the near-infrared photometry of VIKING or UKIDSS and the mid-infrared photometry of unWISE (28 objects). The results indicate that 7 objects exhibit SEDs consistent with galaxies at $z \sim 4$. The HzRG candidates have very large stellar masses with $\sim 4.2 \times 10^{11} M_{\odot}$ on average. This stellar mass is similar to that of previously discovered USS HzRGs at $z \sim 4$, though our sample is affected by a sample selection bias that selects only HzRGs with $M_{\star} > 10^{11} M_{\odot}$. Further, the SEDs of those HzRG
candidates suggest a past fast quenching with a rough timescale of $\sim$0.1 Gyr, as evidenced from the rest-frame UVJ diagram.

\end{abstract}


\section{Introduction} \label{sec:intro}

Various observational studies have reported the positive correlation between the mass of a supermassive black hole (SMBH) and a bulge mass of its host galaxy, suggesting a co-evolutionary relationship between SMBHs and galaxies (e.g., \citealt{1998AJ....115.2285M, 2003ApJ...589L..21M, 2013ARA&A..51..511K}). One proposed model to explain this co-evolution is a scenario wherein two gas-rich galaxies undergo a major merger (e.g., \citealt{1988ApJ...325...74S, 2008ApJS..175..356H}). In this scenario, initially gas-rich galaxies approach due to their gravity, collide, and merge. This process rapidly increases star-formation activity within the galaxy as gas is compressed. Furthermore, gas losing angular momentum begins to accrete towards the SMBH. Consequently the SMBH becomes active and behaves an active galactic nucleus (AGN). The AGN activity generates outward radiation pressure, heating and expelling gas surrounding the AGN, which in turn suppresses the star-formation activity in the galaxy. Eventually, the galaxy is expected to transform into an elliptical or radio galaxy with diminished star-forming activity.

Among AGNs, galaxies that have radio jets and are bright in radio are referred to as radio-loud AGNs. Here the type-2 radio-loud AGN is referred to as the radio galaxy. Radio galaxies are interesting objects from which the stellar component of galaxies can be examined.
Low-redshift ($z < 1$) radio galaxies are characterized by their high mass, with typical stellar mass $\sim 10^{11} M_{\odot}$ (e.g., \citealt{2005MNRAS.362....9B, 2019ApJS..243...15T}). One notable feature of low-$z$ radio galaxies is the suppression of the star formation (e.g., \citealt{2005A&A...444L...9M, 2012MNRAS.421.1569B, 2012ARA&A..50..455F, 2017A&A...600A.121N, 2020ApJ...901..159C}). \cite{2020ApJ...901..159C} reported that radio galaxies exhibit, on average, a star formation rate (SFR) of $\sim$1/30 of that of main-sequence galaxies. These radio galaxies tend to exhibit a low mass-accretion rate onto SMBHs; for example, \cite{2012MNRAS.421.1569B} found that 80\% of radio galaxies have Eddington ratio of $\lambda_{\mathrm{Edd}}<10^{-2}$, assuming that the ratio of the SMBH mass to stellar mass is constant.

High-$z$ ($z > 1$) radio galaxies (HzRGs) are also characterized mostly by the evolved stellar populations with a stellar mass of $\sim 10^{11} M_{\odot}$ (e.g., \citealt{2007ApJS..171..353S, 2010ApJ...725...36D, 2019MNRAS.489.5053S}).
Moreover, radio galaxies exhibit a correlation known as the $K-z$ relation, that is, a tight relation between the redshift and the apparent $K$-band magnitude (e.g., \citealt{1982MNRAS.199.1053L, 2001MNRAS.326.1563J, 2002AJ....123..637D, 2003MNRAS.339..173W, 2010MNRAS.407.1739I}). This correlation is consistent with passively-evolving massive galaxies with a past instantaneous starburst (e.g., \citealt{2003MNRAS.339..173W}). Thus, HzRGs can be considered to reside in environments wherein the structure formation was already advanced in high-redshift Universe, such as proto-clusters of galaxies. HzRGs have been utilized as beacons for identifying and studying proto-clusters and overdensity regions (e.g., \citealt{1996Natur.383...45P, 1996ApJ...471L..11L, 2007A&A...461..823V, 2011MNRAS.410.1537H, 2012ApJ...757...15H, 2014MNRAS.440.3262C, 2022PASJ...74L..27U}). This renders them highly intriguing objects for investigating the nature of HzRGs themselves and their relationship to the environment.

Despite the importance of HzRGs described above, HzRGs with $z > 4$ are rare and challenging to discover (e.g., \citealt{2021A&A...647A...5W, 2023A&A...672A.171I}). An effective method to identify HzRGs is the ultra steep spectrum (USS) technique introduced by \cite{1996ApJS..106..247C}. 
The USS is defined as a radio spectrum characterized by a spectral index $\alpha < -1.3$ where $F_\nu \propto \nu^\alpha$ (e.g., \citealt{2000A&AS..143..303D, 2018MNRAS.480.2733S, 2019MNRAS.489.5053S}). This method facilitated the successful discovery of several HzRGs. However, the number of known HzRGs at $z \gtrsim 4$ remains limited (e.g., \citealt{2008A&ARv..15...67M, 2019MNRAS.489.5053S, 2022PASA...39...61B}). Moreover, this selection method is biased toward HzRGs with the USS feature. Thus, HzRGs without the USS characteristics have been found through alternative selection methods such as the Lyman-break technique \citep{2009MNRAS.398L..83J, 2020AJ....160...60Y, 2023MNRAS.519.4902S}. Hence, there is a need for new explorations that can capture a broader range of HzRGs, including those without significant USS features, to investigate the general population of HzRGs.

To explore HzRGs in a systematic manner, a powerful observational program was conducted recently, which is Hyper Surime-Cam Subaru Strategic Program (HSC-SSP; \citealt{2018PASJ...70S...8A}).
HSC-SSP is an optical multi-band imaging observation that covers a wide area with $\sim$3 mag deeper than the Sloan Digital Sky Survey (SDSS; \citealt{2000AJ....120.1579Y}). \cite{2020AJ....160...60Y} discovered a new HzRG at $z=4.72$ by adopting the $r$-dropout Lyman-break method to the HSC-SSP data, and revealed its massive nature ($M_\star \sim 3 \times 10^{11} M_\odot$). To study the statistical characteristics of HzRGs and their stellar population, this study conducted a new systematic survey of HzRGs using the HSC-SSP data.
Consequently, by selecting HzRGs that are detected in past near-infrared and mid-infrared surveys, we estimated their stellar mass and star formation history. This study focused only on radio galaxies, by excluding radio-loud quasars. This is because the rest-frame UV and optical emission of radio-loud quasars is dominated by power-law AGN radiation, which is not helpful for investigating their host galaxies. Whereas, radio galaxies are type-2 AGNs and thus their rest-frame UV and optical emission are mostly of stellar origin.
This survey is part of an ongoing project called ``A Wide and Deep Exploration of Radio Galaxies with Subaru HSC (WERGS; \citealt{2018ApJ...866..140Y, 2019ApJS..243...15T, 2021ApJ...921...51I})", which is a systematic study of radio galaxies in both low-$z$ and high-$z$ Universe observed by HSC-SSP.

This paper is structured as follows. In Section \ref{sec:ss}, we present the data and analysis methods. Section \ref{sec:results} presents the SED fitting results for HzRG candidates. In Section \ref{sec:disscussion}, we discuss the statistical properties and characteristics of the HzRG candidates. Finally, the conclusion is presented in Section \ref{sec:conclusion}. Throughout the paper, we adopt the cosmological parameters of the $\Lambda$CDM model, specifically $H_{0}=70$ km s$^{-1}$ Mpc$^{-1}$, $\Omega_{M}=0.27$, and $\Omega_{\Lambda}=0.73$, which are the same as those adopted in \cite{2018ApJ...866..140Y}. Unless otherwise stated, all magnitudes are presented in the AB system (\citealt{1983ApJ...266..713O}).

\section{Data and analysis} \label{sec:ss}

\subsection{Data} \label{sec:DC}

In this study, we selected $z\sim4$ radio galaxies based on the following procedure.
The fundamental strategy is the same as the method adopted in \cite{2020AJ....160...60Y}; that is, via the positional matching between a Lyman-break galaxy (LBG) sample selected with the HSC-SSP data and a radio-source catalog made by the Faint Images of the Radio Sky at Twenty cm (FIRST; \citealt{1995ApJ...450..559B, 2015ApJ...801...26H}). HSC \citep{2018PASJ...70S...1M} is a wide-field prime-focus camera with 90 arcmin diameter field-of-view, installed on the Subaru Telescope. HSC employs five optical bands ($g, r, i, z,$ and $y$) for broadband filters (\citealt{2018PASJ...70...66K}). HSC-SSP (\citealt{2018PASJ...70S...8A}) is an optical imaging survey conducted using HSC by spending 330 nights, comprising three layers (ultra-deep, deep, and wide). We utilized the wide-layer data of the S19A data release (\citealt{2022PASJ...74..247A}), covering an observation area of approximately 560 $\mathrm{deg^2}$. The 5-sigma limiting magnitudes for $g, r, i, z,$ and $y$ are 26.5, 26.5, 26.2, 25.2, and 24.4 mag, respectively. The point-spread function (PSF) size in full width at half maximum (FWHM) is $\sim$0.7 arcsec \citep{2024MNRAS.531.2517G}.
In some previous studies on LBGs based on the S16A data release (\citealt{2018PASJ...70S..10O, 2018PASJ...70S..11H, 2022ApJS..259...20H}), the cmodel magnitude that is measured by combining de Vaucouleurs's and exponential profiles \citep{2004AJ....128..502A, 2018PASJ...70S...5B} was used to measure the total fluxes and colors of sources. 
However, \cite{2022ApJS..259...20H} found that the S16A and S19A databases yielded several different values in each other for the cmodel calculations. Consequently, they determined the best agreement between the cmodel of the S16A data release and a 2''-diameter aperture following the aperture correction, that is, convolvedflux\_0\_20\_mag of the S19A data release. The aperture correction factor is calculated in each band by considering the point spread function. In general, the aperture photometry is not suitable for measuring the total flux; however, it can produce consistent values when measuring colors between different bands.
Therefore, the magnitude used in this study is ($g,r,i$)\_convolvedflux\_0\_20\_mag for the LBG election.  

In the later analysis, we also used the cmodel mag, which is the appropriate model magnitude, as the total magnitude.
Each magnitude was corrected for the Galactic extinction (\citealt{1998ApJ...500..525S}). 

FIRST is a 1.4 GHz radio survey obtained with Very Large Array (VLA). The catalog version utilized in this study is the final data release (14Dec17 Version;  \citealt{2015ApJ...801...26H}). The spatial resolution is 5.4 arcsec and the positional accuracy is better than 1.0 arcsec. The limiting flux of FIRST is $\sim$1 mJy at 5 sigma sensitivity. As it covers the entire HSC-SSP observation area, the FIRST survey data are highly suitable for investigating the characteristics of optical sources detected by the HSC in the radio frequency range.

VISTA Kilo-degree Infrared Galaxy Survey (VIKING; \citealt{2007Msngr.127...28A}) is a near-infrared imaging survey conducted using Visible and Infrared Survey Telescope for Astronomy (VISTA), which is  a wide-field near-infrared telescope featuring a 4-meter primary mirror and equipped with a near-infrared camera VIRCAM (VISTA InfraRed CAMera; \citealt{2006SPIE.6269E..0XD}). The 5-sigma limiting magnitudes for $Z$, $Y$, $J$, $H$, and $K_{S}$ are 22.7, 22.0, 21.8, 21.2, and 21.2 mag, respectively. The median PSF size in FWHM is 0.9 arcsec \citep{2014PASA...31....4A}. The data release employed in this study is DR4 (\citealt{2013Msngr.154...32E}). The total observation area of the VIKING survey is 1350 $\mathrm{deg^2}$, with the portion overlapping with the HSC coverage amounting to 262 $\mathrm{deg^2}$.

The UKIRT Infrared Deep Sky Survey (UKIDSS; \citealt{2007MNRAS.379.1599L}) is a near-infrared imaging survey conducted using the United Kingdom Infra-Red Telescope (UKIRT), which is a 3.8-meter telescope dedicated to infrared observations and equipped with the Wide Field Camera (WFCAM; \citealt{2007A&A...467..777C}). The UKIDSS Large Area Survey (LAS) DR11 used in this study was conducted using $Y,J,H$, and $K$ filters with 5-sigma limit magnitudes of 20.8, 20.5, 20.2, and 20.2 mag, respectively. The median PSF size in FWHM is 1.2 arcsec \citep{2018MNRAS.473.5113D}. The total observation area covered by the UKIDSS LAS is 4028 $\mathrm{deg^2}$. For this specific study, a subset of 156 $\mathrm{deg^2}$ within the HSC-SSP observation area was utilized, which is not within the VIKING observation area.

unWISE (\citealt{2019ApJS..240...30S}) is a catalog that combines three-year data of ALLWISE (\citealt{2010AJ....140.1868W}), which is derived from previous WISE observations, and NEOWISE data (\citealt{2014ApJ...792...30M}), which is ongoing observations. The unWISE catalog provides photometric data for two mid-infrared bands at 3.4 ($W1$) and 4.6 ($W2$) $\mu$m. The 5-sigma limit magnitudes for these bands are 20.6 and 20.1 mag, respectively. The median PSF size in FWHM is 6.1 and 6.8 arcsec, respectively \citep{2014AJ....147..108L}. unWISE is an all-sky survey and covers the entire HSC-SSP observation area.

\vspace{1mm}

\subsection{Photometry of VIKING and UKIDSS sources}  \label{subsec:PVU}

This study used the near-infrared VIKING and UKIDSS photometry data for analyzing the spectral energy distribution (SED)  of HzRGs. However, there are some instances wherein an object is visually confirmed in the image while being categorized as undetected in the VIKING and UKIDSS photometric catalogs. Owing to the very low number density of HzRGs, such cases must be addressed. Therefore we relied on photometric information derived directly from the image data instead of the magnitudes listed in the VIKING and UKIDSS photometry catalogs. This approach ensured that the photometric data used for SED fitting of HzRGs included any data that were treated as undetected in the catalogs.
We employed the aperture photometry for $Z$, $Y$, $J$, $H$, and $K_{S}$ in VIKING and $Y$, $J$, $H$, and $K$ in UKIDSS. The photometry center of the target objects was determined using the coordinates from the HSC-SSP data. For each target object, we calculated the sum of pixel counts within an aperture circle with a diameter of 4 arcsec. This specific aperture size was chosen as it almost captures the total flux of the object. To subtract possible background residuals, the average count of the background obtained from the sky area surrounding the source was measured. This sky area was defined as a ring-like region with an inner and outer diameters of 12 and 20 arcsec, respectively, centered on the target source. The count from the sky area was then normalized to an area with a diameter of 4 arcsec and consequently subtracted from the target source counts. 

As described in Section \ref{sec:DC}, photometric data obtained with different instruments exhibit different image quality. To check the consistency of the photometry, we focused on the $y$-band magnitude measured with HSC-SSP, VIKING, and UKIDSS. We randomly selected 1,000 objects that are observed in both HSC-SSP and the near-infrared (VIKING or UKIDSS), and detected with S/N $>$ 10 in the near-infrared. We selected objects whose morphology was neither a point like nor very 
extended\footnote[1]{To quantify the extendedness of HSC sources, we focus on the ratio of the 2nd-order adaptive moment of the source to that of the PSF, measured in the HSC $i$-band. Specifically, we select objects satisfying both $\mathrm{1.1< ishape\_hsm\_moment\_11/ishape\_hsm\_psfmoment\_11<2.0}$ and $\mathrm{1.1<ishape\_hsm\_moment\_22/ishape\_hsm\_psfmoment\_22}$ $<2.0$ to collect objects which are moderately extended. See \cite{2018PASJ...70S..34A} for more details.}. 
As a result, the difference between the HSC and VIKING $y$-band magnitudes, $y_{\rm HSC} - y_{\rm VIKING}$, is --0.02 on average with a standard deviation of 0.22. The difference between the HSC and UKIDSS $y$-band magnitudes, $y_{\rm HSC} - y_{\rm UKIDSS}$, is --0.02 on average with a standard deviation of 0.29. These differences were negligibly small in the following analysis.

To estimate the photometric errors, we employed the following method. We prepared a square image with 12 arcminutes per side, centered on each target object. Subsequently, the aperture photometry for 10,000 random coordinates was performed by adopting the same method as for the target objects. Since some of the 10,000 photometric measurements were affected by astronomical objects, we created a 3 sigma clipping for the photometric measurements and derived the standard deviation to estimate the sky fluctuation that corresponded to the photometric error for the target objects.

\subsection{Selection of HzRGs} \label{subsec:Hs}

\subsubsection{LBG selection} \label{subsubsec:LBG}

We first created a clean sample of HSC sources by excluding false objects in the sample and objects with unreliable photometric values. The flags used for creating the clean sample are summarized in Table \ref{table:data_hsc}. In addition, we adopted a criterion of $i$-band signal-to-noise ratio (S/N) $>$ 5 for secure detections of sources.

\begin{table*}[t]

\begin{center}
\caption{Criteria to make the HSC clean sample}
\label{table:data_hsc}
\begin{tabular}{lccl} \hline
   Parameter & Value & Band & Comment \\ \hline \hline

detect\_is\_primary&True&---&Object is a primary one with no deblended children. \\
pixelflags\_edge&False&$g, r, i, z, y$&Located within images. \\
pixelflags\_saturatedcenter &False&$g, r, i, z, y$&None of the central $3 \times 3$ pixels of an object is saturated. \\
pixelflags\_bad&False&$g, r, i, z, y$&None of the pixels in the footprint of an object is labelled as bad. \\
sdsscentroid\_flag&False&$r, i$&Object centroid measurement has no problem. \\
cmodel\_flux\_flag&False&$g, r, i$& Cmodel flux measurement has no problem. \\
mask\_s18a\_bright\_objectcenter&False&$g, r, i, z, y$&None of the pixels in the footprint of an object is close to bright sources. \\
merge\_peak&True&$r, i$&5 $\sigma$ detected in $r$ and $i$ and peak position of $r$ and $i$ is within 3''.\\
blendedness\_abs&$\leq$0.2&$r, i$& The target photometry is not significantly affected by neighbors. \\
inputcount\_value&$\geq$ (3, 3, 5, 5, 5)&$g, r, i, z, y$& Number of images contributing at center. \\

    \hline
\end{tabular}

\end{center}

\end{table*}

\begin{figure}
 \begin{center}
 \includegraphics[width=9.0cm]{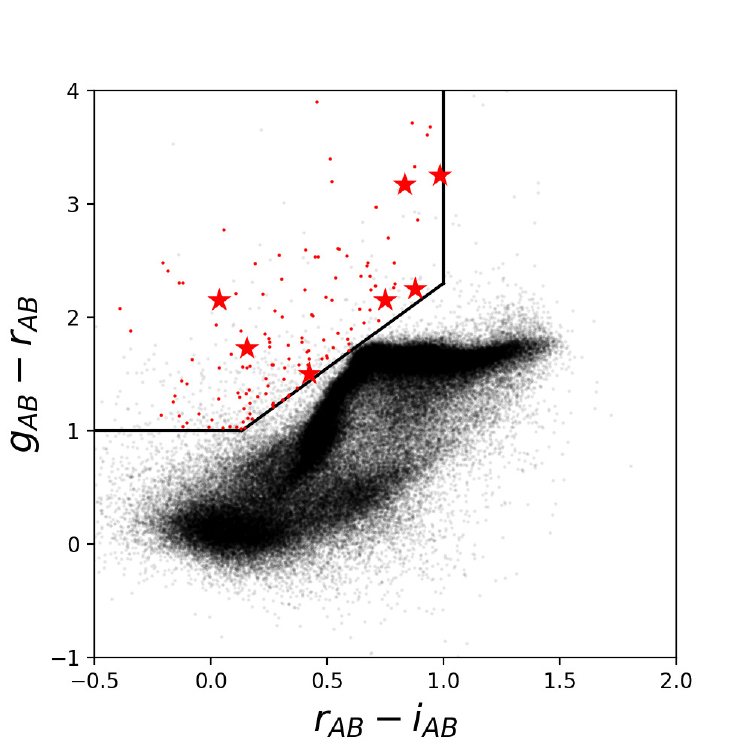}
 \end{center}
 \caption{
 Distribution of $g-r$ and $r-i$ colors. The black dots denote sources with spectroscopic redshift within the range of $0.1<z_{\rm sp}<3.0$, provided in the spectroscopic catalog of HSC-SSP S19A. The red dots show the 146 HzRG candidates, and the red stars denote the 7 finally selected HzRG candidates for the SED analysis (see Section \ref{subsec:SEDA}).}
\label{fig:gri}
\end{figure}

The $g$-dropout LBGs were selected from the clean sample by applying the color criteria shown below (Figure \ref{fig:gri}), which were the same as the criteria adopted by the earlier HSC-SSP studies on LBGs  \citep{2018PASJ...70S..10O, 2018PASJ...70S..11H, 2022ApJS..259...20H}. 
\begin{quote}
 \begin{itemize}
  \item $g$ $-$ $r$ $\textgreater$ 1.0
  \item $r$ $-$ $i$ $\textless$ 1.0
  \item $g$ $-$ $r$ $\textgreater$ 1.5 $\times$ ($r$ $-$ $i$) $+$ 0.8
 \end{itemize}
\end{quote}
The number of $g$-dropout LBGs selected through these criteria was 2,553,430 in $\sim$560 $\mathrm{deg^2}$.

\citet{2018PASJ...70S..10O} reported the expected redshift range of the $g$-dropout LBGs selected through these color criteria with the HSC photometric system to be approximately $3.3<z<4.5$. They also showed the distribution of the spectroscopic redshift of $g$-dropout LBGs selected with HSC-SSP, which is consistent to the expected redshift range. 

\subsubsection{Cross matching of the LBG sample and radio catalog} \label{subsubsec:Cm}

We selected photometric candidates of HzRGs at $z \sim 4$ by matching the $g$-dropout LBG sample with the coordinates of the FIRST radio sources. The search radius to match the LBG sample and the FIRST sample was 1 arcsec, as adopted by \cite{2018ApJ...866..140Y} who examined the optimal search radius to match the HSC-SSP and FIRST samples. \cite{2018ApJ...866..140Y} reported that, through the adoption of the matching radius of 1 arcsec, the contamination by chance coincidence was estimated to be 14\%. Further, the completeness was estimated to be 93\%.
Thus, by adopting a matching radius of 1 arcsec in our study, we aimed to explore the radio emission from LBGs, allowing us to identify HzRGs at $z \sim 4$. 
Through this matching procedure, we obtained 184 objects. Notably, among these 184 objects, there were no other HSC sources within 1 arcsec from the FIRST positions.

\subsubsection{Contamination} \label{subsubsec:Contami}

There is a possibility that low-$z$ radio galaxies and high-$z$ radio-loud quasars are also present among the HzRG candidates. To mitigate this, we excluded as many of such contaminations as possible. Specifically, we investigated whether any of the 184 HzRG candidates had already undergone spectroscopic observation and had known redshifts and photometric characteristics. For this purpose, we referred to the spectroscopic catalogs available in the HSC-SSP S19A database.

\begin{figure}
 \begin{center}
 \includegraphics[width=9.0cm]{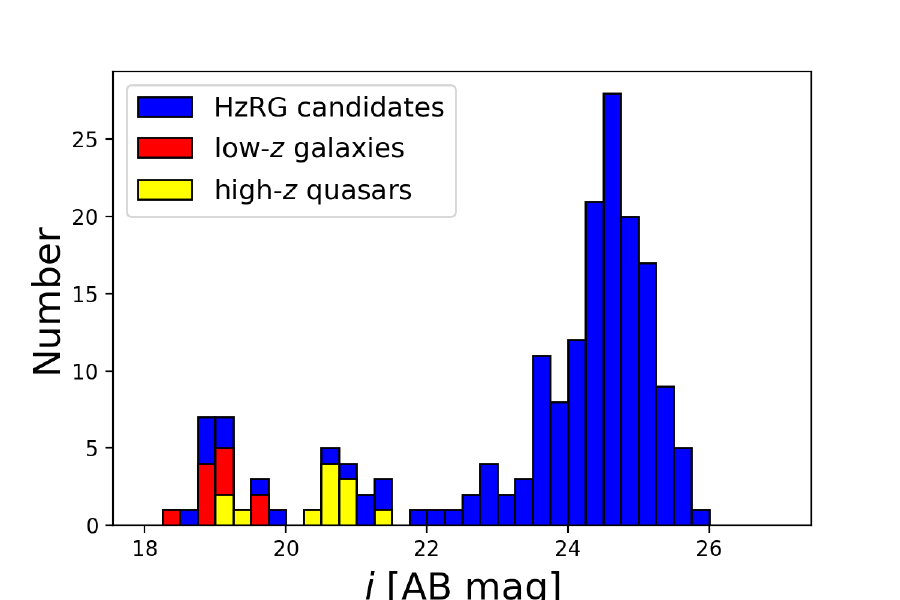}
 \end{center}
 \caption{Distribution of $i$-band magnitude for 184 HzRG candidates. Blue histogram shows all the 184 HzRG candidates, red shows 11 low-$z$ galaxies selected as HzRG candidates, and yellow shows the 12 spectroscopically-confirmed high-$z$ quasars.}
\label{fig:BUMP_PL_hist}
\end{figure}

We searched spectroscopic data of the 184 HzRG candidates. Consequently, 23 were found to have spectroscopic information.
We show the distribution of $i$-band magnitudes for 184 HzRG candidates and 23 objects having spectroscopic observations in Figure \ref{fig:BUMP_PL_hist}. Among these, 11 low-$z$ galaxies ($z<1.0$) and 12 high-$z$ quasars ($3.0<z<5.0$) were identified. Figure \ref{fig:BUMP_PL_hist} shows that the low-$z$ galaxies and high-$z$ quasars in our sample of HzRG candidates had relatively bright $i$-band magnitudes (brighter than 21.5 mag). As more than half of the HzRG candidates with $i<21.5$ were such contaminations (low-$z$ galaxies or high-$z$ quasars), we considered only $g$-dropout objects with $i>21.5$ as candidates of HzRGs hereafter. As a result, the number of HzRG candidates was reduced to 146. However, this sample of HzRG candidates (with $i > 21.5$) can be still contaminated by low-$z$ galaxies and high-$z$ quasars. Thus, we selected the most reliable HzRGs among the candidates through the SED analysis (details in Sections \ref{subsec:SEDA} and \ref{sec:results}).

\subsubsection{Multi-band information} \label{subsubsec:Mbi}

To conduct the SED analysis of the HzRG candidates, we integrated data from the optical observations of HSC-SSP and the infrared datasets of VIKING$+$UKIDSS and unWISE. First, we obtained unWISE mid-infrared data for 75 objects among the 146 HzRG candidates. These 75 sources were detected in $W1$ and/or $W2$ bands with $>$5 $\sigma$, whereas the other 71 sources were not detected in both the $W1$ and $W2$ bands. This was achieved by performing a positional match with a search radius of 3 arcsec with the unWISE catalog, adopting the same search radius as adopted by \citet{2019ApJS..243...15T}. Second, we acquired VIKING data for 21 sources, and UKIDSS data for an additional 7 sources.  Among these 28 sources, the near-infrared photometric magnitudes of 17 objects are provided in the VIKING (16 objects) or UKIDSS (1 object) catalogs, and magnitudes of 11 objects are not presented in the two catalogs. Therefore our own photometry described in Section \ref{subsec:PVU} was actually effective to search for HzRGs. In the analysis given below, we adopted our own photometry for near-infrared magnitudes for all 28 sources. By combining these mid- and near-infrared photometric data, we can study the SED in the range of $0.5 \ \mu{\rm m} \lesssim \lambda_{\rm obs} \lesssim 4.6 \ \mu{\rm m}$.

\subsection{SED analysis} \label{subsec:SEDA}

We removed possible contaminants from the HzRG candidates to create a reliable sample of HzRGs at $z \sim 4$. For this purpose, we conducted SED fitting using a tool called X-CIGALE (\citealt{2020MNRAS.491..740Y}; see also \citealt{2005MNRAS.360.1413B, 2009A&A...507.1793N, 2019A&A...622A.103B}) to estimate the photo-$z$ of the HzRG candidates. X-CIGALE can perform photo-$z$ estimation (e.g., \citealt{2014A&A...562A..15M, 2020ApJ...899...35T}). X-CIGALE is designed for SED modeling, by ensuring the energy conservation in the UV/optical and infrared regimes. For the SED fitting, we employed a galaxy model that considers star formation history, stellar population, nebular emission, dust attenuation, and redshift as parameters (Table \ref{table:XCIGALE_para1}). As the available photometric information was not sufficient to precisely determine detailed star formation history, a simple star formation model known as the delayed model was adopted. It is defined as: $$\mathrm{SFR} \ (t) \propto \frac{t}{\tau} \times \mathrm{exp} \  (- \ \frac{t}{\tau}),$$ where $t$ is the time after the start of the star formation, and $\tau$ is the epoch of the SFR peak. The stellar population model used in the SED fitting was BC03 (\citealt{2003MNRAS.344.1000B}), and we assumed the initial mass function (IMF) of \citet{2003PASP..115..763C}. The stellar metallicity was assumed to be the Solar value ($Z = 0.02$). The nebular emission module, based on the empirical templates of \citet{2011MNRAS.415.2920I}, describes the line emission, which is sometimes essential for strong emission-line galaxies \citep{2013ApJ...763..129S}. The ionization parameter $U$ is assumed to be --4.0 or --2.0. To model the dust attenuation, the extinction law of \citet{2000ApJ...533..682C} was adopted. $E(B-V)$ was parameterized from 0 to 2 to reproduce dusty objects as well (see Table \ref{table:XCIGALE_para1}). In the fitting process, the data were treated as 2 sigma upper limits if their detection significance was below 2 sigma for HSC-SSP, UKIDSS, and VIKING data, and as 5 sigma upper limits for unWISE data if the detection was below 5 sigma. We performed the SED fitting for a total of 28 sources: 21 HzRG candidates with VIKING and 7 HzRG candidates with UKIDSS. Because no objects among the 28 HzRG candidates exhibited a spectroscopic redshift, the photometric redshift was also estimated during the SED fitting, as described in Section \ref{subsec:SEDp}.

We did not include the AGN component in the SED fit with X-CIGALE. This is because the continuum emission in the rest-frame UV and optical range of radio galaxies is mostly dominated by the stellar component owing to the type-2 AGN nature of radio galaxies. Although the AGN emission is significant in the rest-frame infrared wavelength, it is beyond the scope of this study.

\begin{table}[h]
  \caption{Model parameters for X-CIGALE fits}
  \label{table:XCIGALE_para1}
  \centering
  \begin{tabular}{|l|l|}
    \hline
      \multicolumn{2}{|c|}{Star formation history (SFH)} \\
    \hline \hline
    Model  & sfhdelayed \\
    $\tau$ (Myr) & 10, 20, 50, 100, 200, 500, \\
      & 1000, 2000 \\
      $\mathrm{age}$ (Myr) & 100, 200, 300, 400, 500, 600, \\
      & 700, 800, 900, 1000, 2000, 4000, \\
      & 8000, 13000 \\
    \hline
      \multicolumn{2}{|c|}{SSP} \\
     \hline \hline
     Model & Bruzual and Charlot (2003) \\
     IMF & Chabrier (2003) \\
     Metallicity & 0.02 \\
     \hline
      \multicolumn{2}{|c|}{Nebular emission} \\
     \hline \hline
     $\mathrm{log}\  U$ & $-4.0, -2.0$ \\
     \hline
      \multicolumn{2}{|c|}{Dust attenuation} \\
     \hline \hline
     Model & Calzetti et al. (2000) \\
     $E(B-V)_{\mathrm{lines}}$ &  0.01, 0.1, 0.2, 0.4, 0.6, 0.8, \\
      & 1.0, 1.2, 1.4, 1.6, 1.8, 2.0 \\
      \hline
  \end{tabular}
\end{table}


\section{Results} \label{sec:results}

\subsection{Photometric redshift} \label{ssubec:Pred}
\begin{figure}
 \begin{center}
 \includegraphics[width=9.0cm]{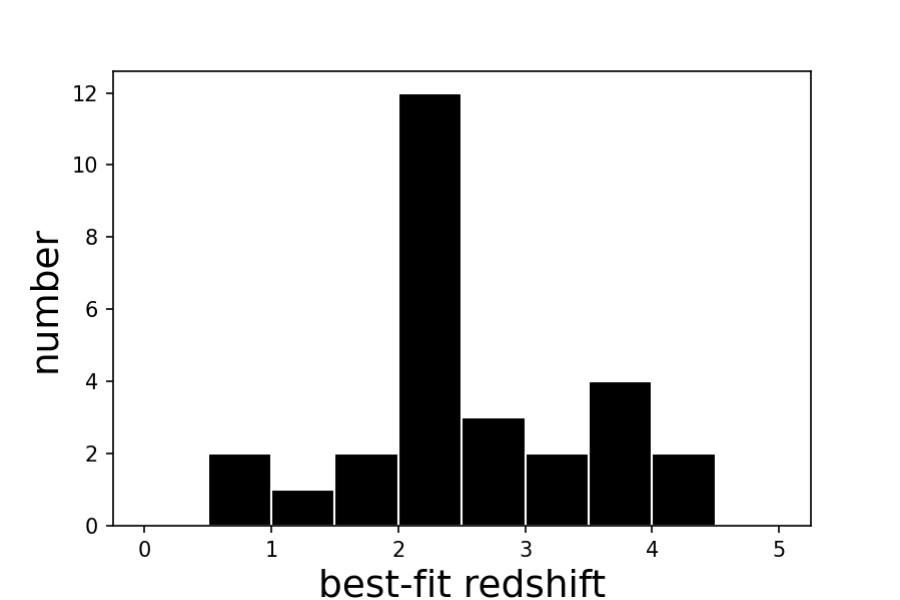}
 \end{center}
 \caption{Best-fit redshift distribution of 28 HzRG candidates.}
\label{fig:z_hist}
\end{figure}

  \begin{figure*}[htbp]
  \begin{center}
   \includegraphics[width=19cm,angle=0]{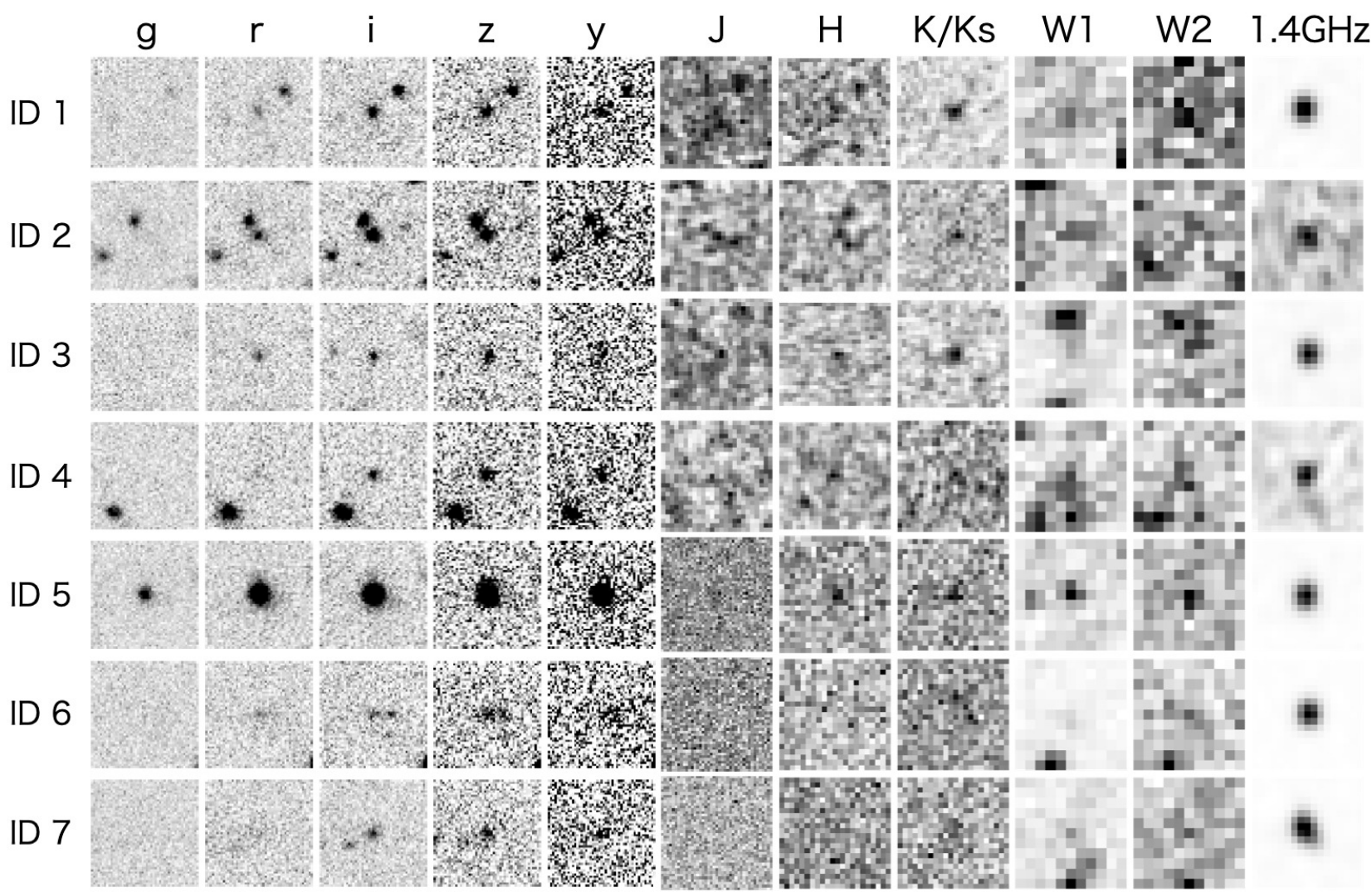}
  \caption{Images of the 7 HzRG candidates of HSC-SSP, VIKING or UKIDSS, unWISE, and FIRST, from left to right. The HSC-SSP $grizy$ and VIKING $JHK_{S}$ or UKIDSS $JHK$ images are shown with a width of 8$^{\prime\prime}$, and subsequently the WISE W1W2 and FIRST 1.4 GHz images with a width of 20$^{\prime\prime}$. North is upward and east is to the left.}
  \label{fig:ID14}
 \end{center}
\end{figure*}

\begin{figure*}[htbp]

  \begin{tabular}{cc}
  
    \begin{minipage}[t]{0.45\linewidth}
      \centering
      \includegraphics[keepaspectratio, width=80mm]
      {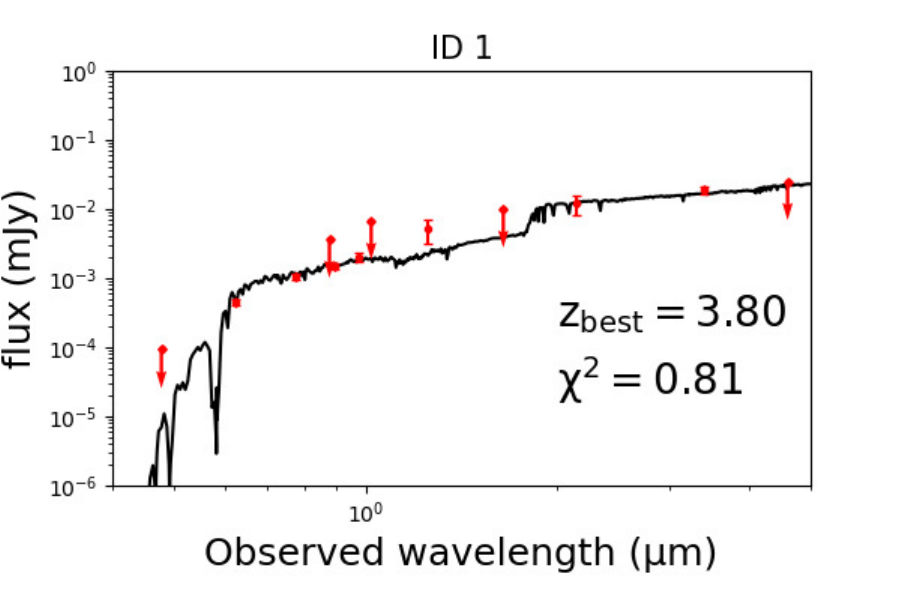}
    \end{minipage} &
    
    \begin{minipage}[t]{0.45\linewidth}
      \centering
      \includegraphics[keepaspectratio, width=80mm]
      {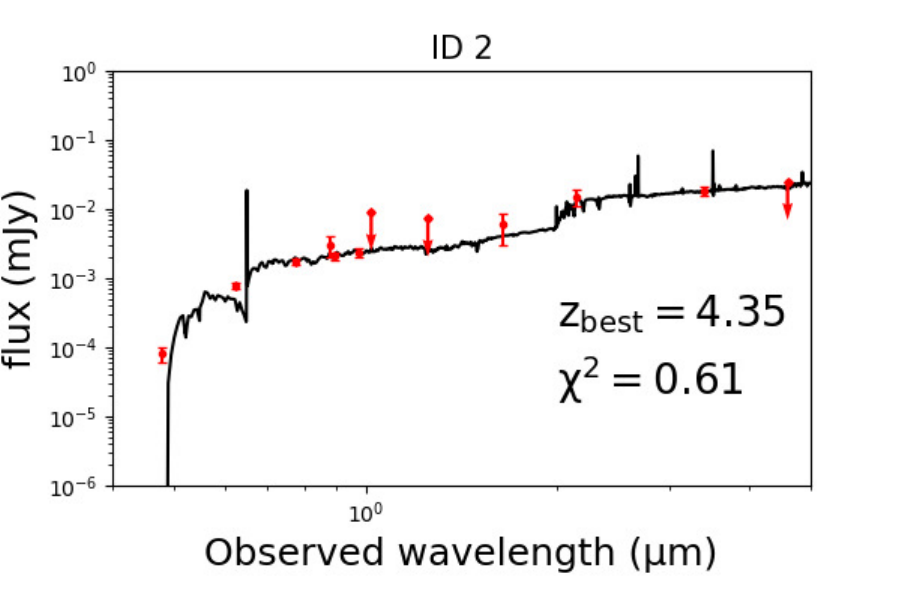}
      \end{minipage} \\

    \begin{minipage}[t]{0.45\linewidth}
      \centering
      \includegraphics[keepaspectratio, width=80mm]
      {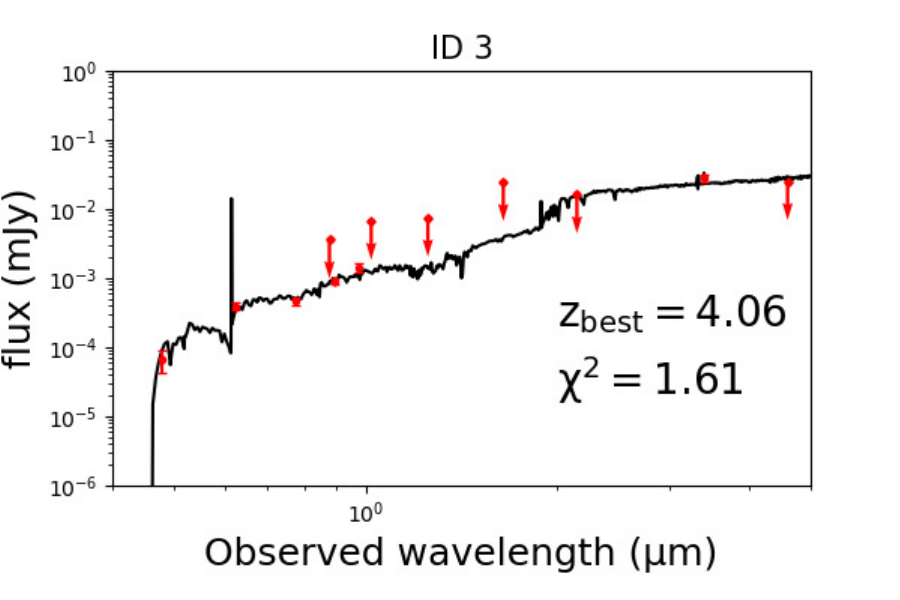}
    \end{minipage}  &
    
    \begin{minipage}[t]{0.45\linewidth}
      \centering
      \includegraphics[keepaspectratio, width=80mm]
      {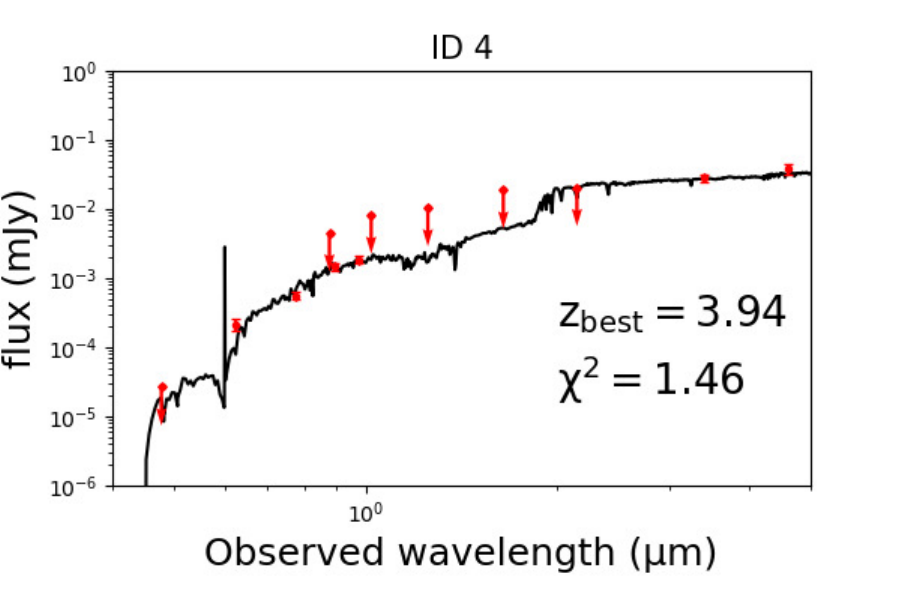}
      \end{minipage} \\
      
       \begin{minipage}[t]{0.45\linewidth}
      \centering
      \includegraphics[keepaspectratio, width=80mm]
      {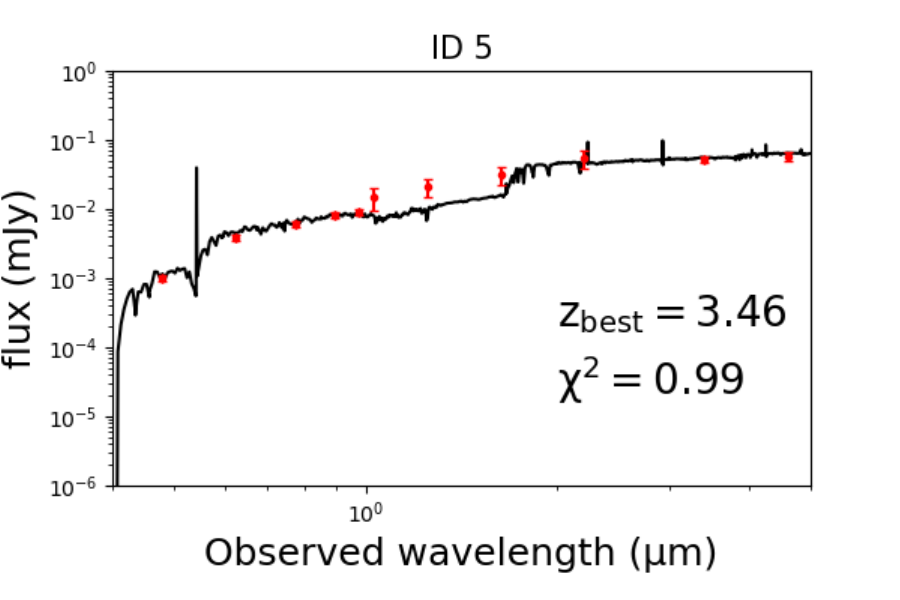}
      
      \end{minipage}  &

    \begin{minipage}[t]{0.45\linewidth}
      \centering
      \includegraphics[keepaspectratio, width=80mm]
      {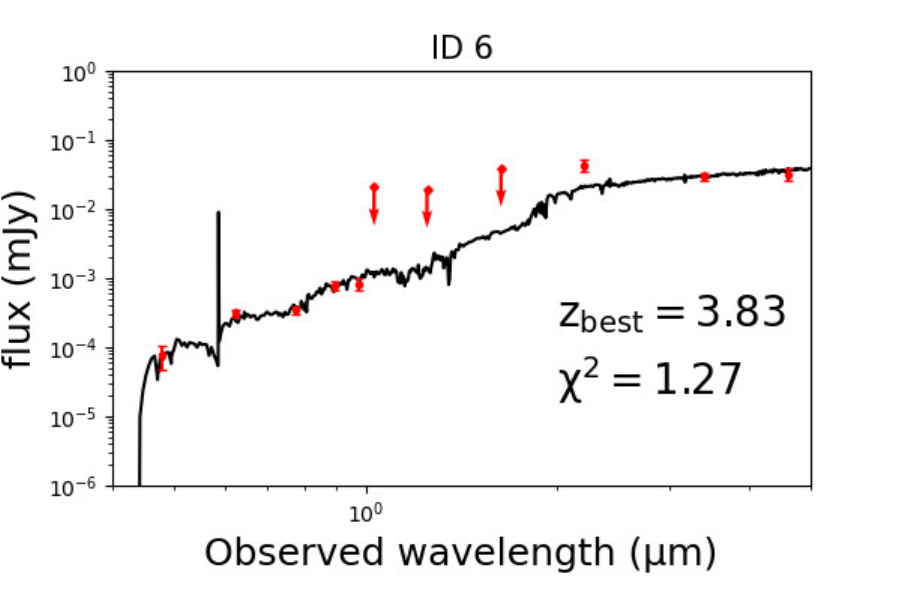}
      \end{minipage}  \\
      
       \begin{minipage}[t]{0.45\linewidth}
      \centering
      \includegraphics[keepaspectratio, width=80mm]
      {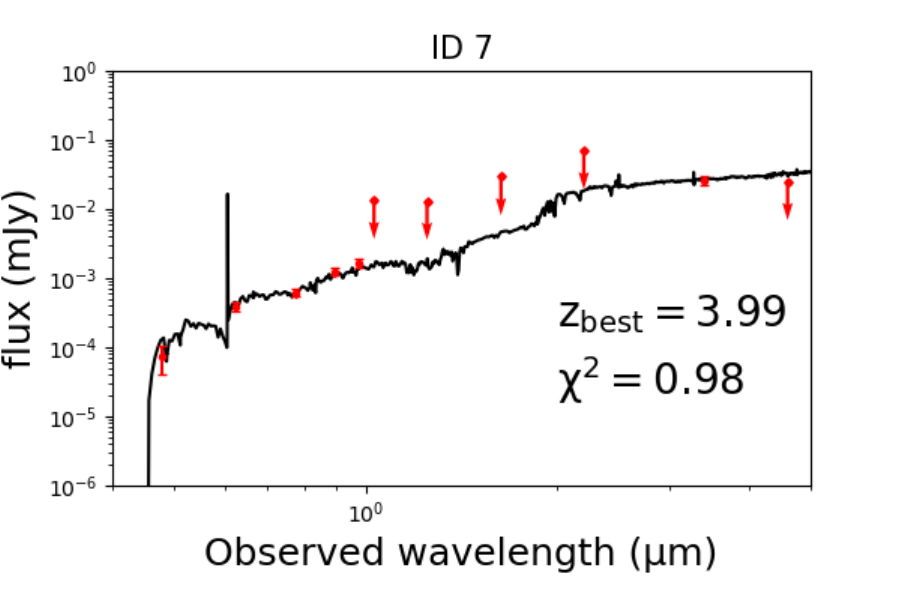}


    \end{minipage} 
  \end{tabular}
  \caption{SED fitting results for the 7 HzRG candidates. Black lines represent the best-fit SED models. Filled red circles denote the photometric data measured with HSC, VIKING (ID 1, 2, 3, and 4) or UKIDSS (ID 5, 6, and 7), and unWISE, while red arrows denote the 2 $\sigma$ upper limit in HSC, VIKING, and UKIDSS and the 5 $\sigma$ upper limit in unWISE. The photo-$z$ and the reduced $\chi^2$ of the best-fit model for each object are shown at the lower-right side in each panel.} 
  \label{fig:sedplot}
\end{figure*}

\begin{figure}
 \begin{center}
 \includegraphics[width=8.5cm]{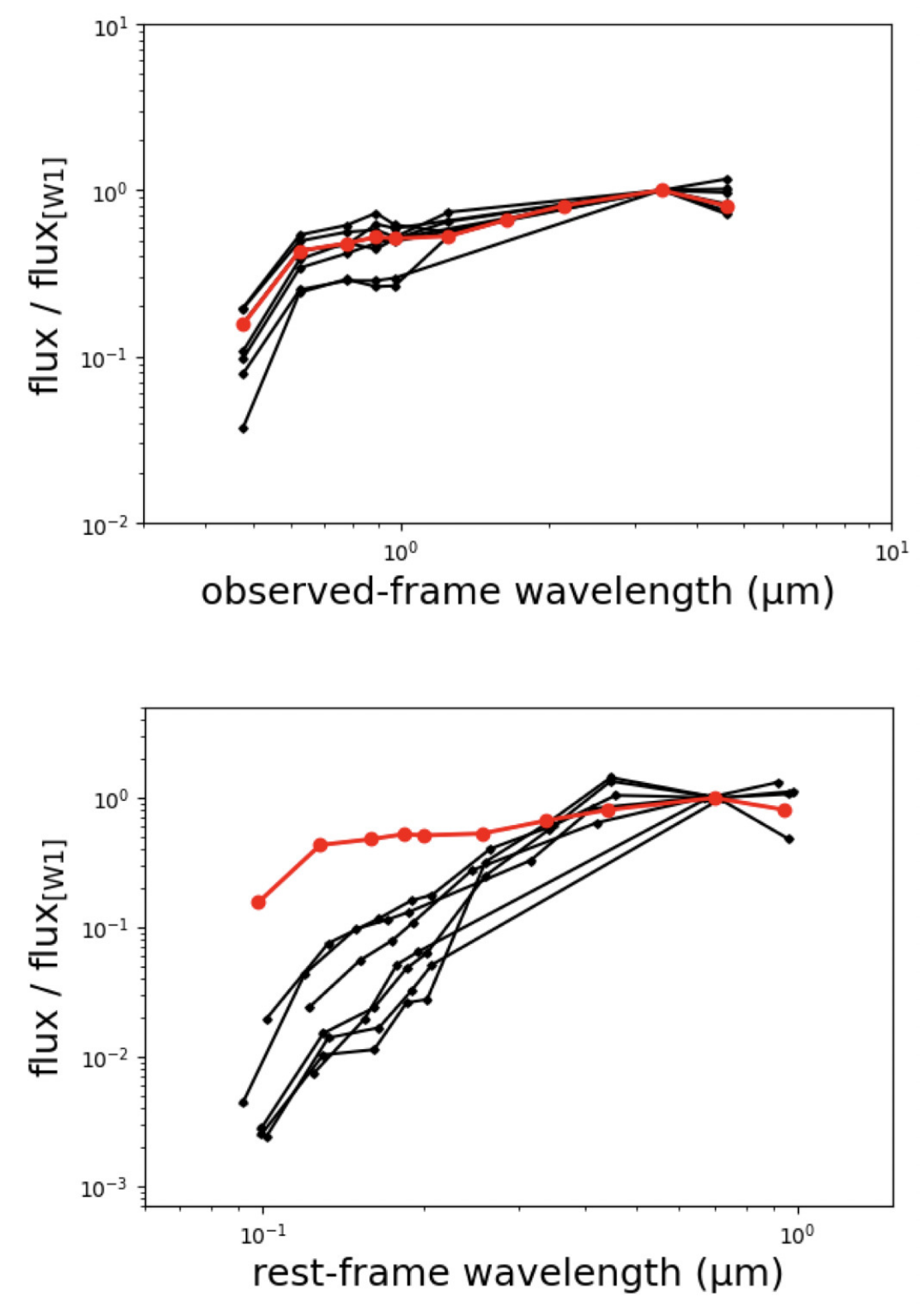}
 \end{center}
 \caption{
 The upper panel shows the observed-frame SED of the 9 high-$z$ quasars (black) and the median SED of those 9 high-$z$ quasars (red) with the photometric data of HSC, VIKING or UKIDSS, and unWISE, normalized by the unWISE W1 flux. The lower panel compares the rest-frame SED of the 7 HzRG candidates (black) and the median SED of high-$z$ quasars (red), normalized by the unWISE W1 flux. 
 }
\label{fig:hzqso}
\end{figure}

Figure \ref{fig:z_hist} shows the distribution of the best-fit photometric redshift of the 28 HzRG candidates selected in Section \ref{sec:ss}. The distribution was considerably wider than the expected redshift range; its median, average, and standard deviation were 2.30, 2.54, and 1.00. This wide distribution of the photometric redshift was attributed to the contamination of low-redshift galaxies ($\sim$ 70\% of the parent sample). Notably, many of those contaminating galaxies were at $z \sim 2$. \cite{2024MNRAS.527..403K} reported that fast-quenching galaxies at $z \sim 2$ exhibited a spectral break at $\lambda_{\rm rest} \sim 1600 \ \mathrm{\AA}$, which can be misidentified as the bright near-infrared feature of $z \sim 4$ galaxies. 
We selected objects exhibiting the photometric redshift in the range of $3.3 < z_{\rm ph} < 4.5$ as the final HzRG candidates, based on the SED fit with X-CIGALE (Sections \ref{subsubsec:LBG} and \ref{subsec:SEDA}). This redshift range was expected for $g$-dropout LBGs selected through the HSC broad-band photometry (Section \ref{subsubsec:LBG}). Consequently, we selected 7 objects as the final HzRG candidate sample. The best-fit redshift of the final sample of the 7 HzRG candidates is summarized in Table \ref{table:data_photoz}. The images of the 7 HzRG candidates are shown in Figure \ref{fig:ID14}. The photometric information for the 7 HzRG candidates is shown in Table \ref{table:data_mag}. The best-fit models well described the observed photometric data, as the reduced $\chi^2$ of the best-fit models was moderately small ($\sim$0.6--1.6). The median, average, and standard deviation of the best-fit redshift were 3.93, 3.91, and 0.27, respectively. The SED plotting the observed photometric data with the best-fit model for the 7 HzRG candidates is shown in Figure \ref{fig:sedplot}. 
The SED of the 7 HzRG candidates are compared with the SED of high-$z$ radio-loud quasars in our original sample (Section \ref{subsubsec:Contami}) in Figure \ref{fig:hzqso}. Among the 12 high-$z$ quasars in our original sample, 9 objects have VIKING or UKIDSS near-infrared data and unWISE mid-infrared data. We obtained the median SED of the 9 high-$z$ quasars in the observed frame (the upper panel in Figure \ref{fig:hzqso}). It is compared with the SED of the 7 HzRG candidates in the rest frame by assuming the median spec-$z$ of the 9 quasars ($z_{\rm sp} = 3.87$) in the lower panel of Figure \ref{fig:hzqso}.
The median SED of high-$z$ quasars is consistent to the power-law continuum emission at the longer wavelength side than the Lyman-break feature, and significantly bluer than our HzRG candidates. This suggest that the radio-loud quasars were successfully removed by the bright-magnitude cut (Section \ref{subsubsec:Contami}).

\begin{table*}[t]

\begin{center}
\caption{Best-fit redshift and reduced $\chi^2$ of the 7 HzRG candidates}
\label{table:data_photoz}
\begin{tabular}{cccc} \hline

   $\mathrm{ID^{1}}$ & Name & Best-fit.Redshift & $\mathrm{reduced \ \chi^{2}}$  \\ \hline \hline
1&HSCJ145117.98$+$005348.7&$3.80$&0.81 \\
2&HSCJ145153.78$+$004152.7&$4.35$&0.61 \\
3&HSCJ122115.75$+$001225.1&$4.06$&1.61 \\
4&HSCJ092207.70$-$014006.2&$3.94$&1.46 \\
5&HSCJ085505.52$+$034804.9&$3.46$&0.99 \\
6&HSCJ234821.76$-$001923.0&$3.83$&1.27 \\
7&HSCJ093820.82$-$004333.4&$3.99$&0.98 \\

    \hline
\end{tabular}
\\ ${}^{\mathrm{1}}$ ID 1, 2, 3, and 4 use VIKING as near-infrared data, while ID 5, 6, and 7 use UKIDSS as near-infrared data.
\end{center}

\end{table*}

\begin{table*}[t]

\begin{center}
\caption{Magnitude information for ID 1, 2, 3, and 4 with HSC, VIKING, and unWISE data, and ID 5, 6, and 7 with HSC, UKIDSS, unWISE data$^1$. The flux information of FIRST is given in unit of mJy, at the lower-right end. }
\label{table:data_mag}
\begin{tabular}{ccccccccccccc} \hline

   ID & $g$ & $r$ & $i$ & $z$ & $y$ & $Z$ & $Y$  \\ \hline \hline
1&$> 27.49$&$24.87\pm0.08$&$23.93\pm0.04$&$23.53\pm0.06$&$23.18\pm0.13$&$> 23.99$&$> 22.75$\\
2&$26.79\pm0.29$&$24.26\pm0.04$&$23.36\pm0.02$&$23.16\pm0.04$&$23.01\pm0.09$&$22.76\pm0.38$&$> 23.02$ \\
3&$26.92\pm0.41$&$24.98\pm0.07$&$24.77\pm0.07$&$24.04\pm0.07$&$23.53\pm0.12$&$> 24.01$&$> 23.34$ \\
4&$> 27.49$&$25.66\pm0.18$&$24.58\pm0.05$&$23.52\pm0.06$&$23.27\pm0.08$&$> 23.76$&$> 23.12$ \\
5&$24.03\pm0.02$&$22.53\pm0.01$&$22.02\pm0.01$&$21.66\pm0.01$&$21.54\pm0.02$&---&$21.01\pm0.40$\\
6&$26.81\pm0.42$&$25.24\pm0.11$&$25.12\pm0.10$&$24.20\pm0.12$&$24.13\pm0.19$&---&$> 21.35$ \\
7&$26.86\pm0.51$&$24.99\pm0.15$&$24.47\pm0.05$&$23.70\pm0.08$&$23.40\pm0.13$&---&$> 21.83$ \\
\hline
ID & $J$ & $H$ & $Ks$ & $K$ & W1 & W2 & FIRST flux \\ \hline \hline
1&$22.16\pm0.41$&$> 22.16$&$21.22\pm0.35$&---&$20.72\pm0.11$&$> 20.10$ & $16.60\pm0.14$ \\
2&$> 22.92$&$> 21.99\pm0.53$&$20.98\pm0.28$&---&$20.75\pm0.13$&$> 20.10$ & $1.08\pm0.15$ \\
3&$> 23.22$&$> 23.22$&$> 21.41$&---&$20.28\pm0.08$&$> 20.10$ & $12.71\pm0.16$ \\
4&$> 22.85$&$> 22.85$&$> 21.53$&---&$20.25\pm0.08$&$19.95\pm0.14$ & $1.22\pm0.15$ \\
5&$20.64\pm0.29$&$20.18\pm0.29$&---&$19.58\pm0.30$&$19.62\pm0.04$&$19.49\pm0.09$ & $21.39\pm0.14$ \\
6&$> 21.44$&$> 20.66$&---&$19.82\pm0.20$&$20.20\pm0.08$&$20.13\pm0.17$ & $23.82\pm0.10$ \\
7&$> 21.86$&$> 20.97$&---&$> 20.04$&$20.36\pm0.09$&$> 20.10$ & $33.16\pm0.15$ \\
\hline

\end{tabular}
\\ ${}^{\mathrm{1}}$ The limiting magnitudes are 2 sigma for HSC, VIKING and UKIDSS, and 5 sigma for unWISE.

\end{center}

\end{table*}

In X-CIGALE, the best-fit model is selected based on the smallest reduced $\chi ^ 2$. However, parameter errors are not evaluated for best-fit models, rather, they are for Bayesian estimates of model parameters. One caveat to evaluate the Bayesian model parameters is that they can significantly deviate from the best-fit estimates when the probability distribution of the photometric redshift exhibits double (or multiple) peaks. This is because the Bayesian model parameters and their errors are evaluated by considering the entire probability distribution of the photometric redshift. To avoid such cases, we performed the second X-CIGALE runs for the final HzRG candidate sample by limiting the search range of the photometric redshift within 3.3 -- 4.5. Based on this limitation, we avoided multiple peaks of the redshift probability distribution and focused only on the peak around the best-fit model.

\subsection{SED parameters} \label{subsec:SEDp}

The evaluated parameters of the best-fit models (Table \ref{table:data_photoz}) are summarized in Table \ref{table:data_bestfit}, and the Bayesian estimates of the dust extinction and stellar mass with the estimated errors are presented in  Table \ref{table:data_bayesfit}. 
The best-fit redshift and Bayesian redshift were nearly consistent within the range of the estimated uncertainty, although the discrepancy was moderately large for ID 5. This relatively large discrepancy is likely because the probability distribution function of the redshift and $E(B-V)$ of ID 5 is broad and asymmetric. Among the model parameters, the age of the stellar population was allowed to be up to 1.3 Gyr, which corresponds to the age of the Universe at $z \sim 4$.

\begin{table*}[t]

\begin{center}
\caption{Best-fit parameters of the 7 HzRG candidates}
\label{table:data_bestfit}
\begin{tabular}{cccccc} \hline

  ID & Best.Redshift & Best.$\tau$ (Myr) & Best.$\mathrm{age}$ (Myr) & Best.$\mathrm{log}\  U$ & Best.$E(B-V)_{\mathrm{lines}}$ \\ \hline \hline
1&$3.80$&$10$&$200$ & $-4.0$ &$0.40$ \\
2&$4.35$&$100$&$600$ & $-2.0$ &$0.20$ \\
3&$4.06$&$100$&$900$ & $-4.0$ &$0.01$ \\
4&$3.94$&$50$&$600$ & $-2.0$ &$0.01$ \\
5&$3.46$&$50$&$400$ & $-4.0$ &$0.10$ \\
6&$3.83$&$100$&$1000$ & $-2.0$ &$0.01$ \\
7&$3.99$&$100$&$900$ & $-2.0$ &$0.01$ \\

    \hline
\end{tabular}

\end{center}

\end{table*}


\begin{table*}[t]
\centering
\begin{center}
\caption{Bayesian estimates of model parameters of the 7 HzRG candidates}
\label{table:data_bayesfit}
\begin{tabular}{cccc} \hline

   ID & Bayes.Redshift & Bayes.$E(B-V)_{\mathrm{lines}}$ & Bayes.$M_{\mathrm{star}} \ [M_{\odot}]$ \\ \hline \hline

1&$4.07\pm0.29$&$0.6\pm0.3$&$(2.8 \pm 1.2) \times 10^{11}$ \\
2&$4.16\pm0.24$&$0.5\pm0.3$&$(2.6 \pm 0.9) \times 10^{11}$ \\
3&$3.74\pm0.28$&$0.6\pm0.6$&$(4.3 \pm 1.2) \times 10^{11}$ \\
4&$4.03\pm0.25$&$0.6\pm0.6$&$(5.2 \pm 1.7) \times 10^{11}$ \\
5&$3.71\pm0.16$&$0.5\pm0.3$&$(4.7 \pm 1.6) \times 10^{11}$ \\
6&$3.80\pm0.17$&$0.1\pm0.1$&$(5.6 \pm 0.8) \times 10^{11}$ \\
7&$3.83\pm0.24$&$0.6\pm0.5$&$(4.6 \pm 1.3) \times 10^{11}$ \\

    \hline
\end{tabular}

\end{center}

\end{table*}


\section{Discussion} \label{sec:disscussion}

\subsection{Selection effects} \label{subsec:Se}
\begin{figure*}[htbp]
\centering
  \begin{tabular}{ccc}
  
    \begin{minipage}[t]{0.33\linewidth}
      \centering 
      \includegraphics[keepaspectratio, width=61mm]{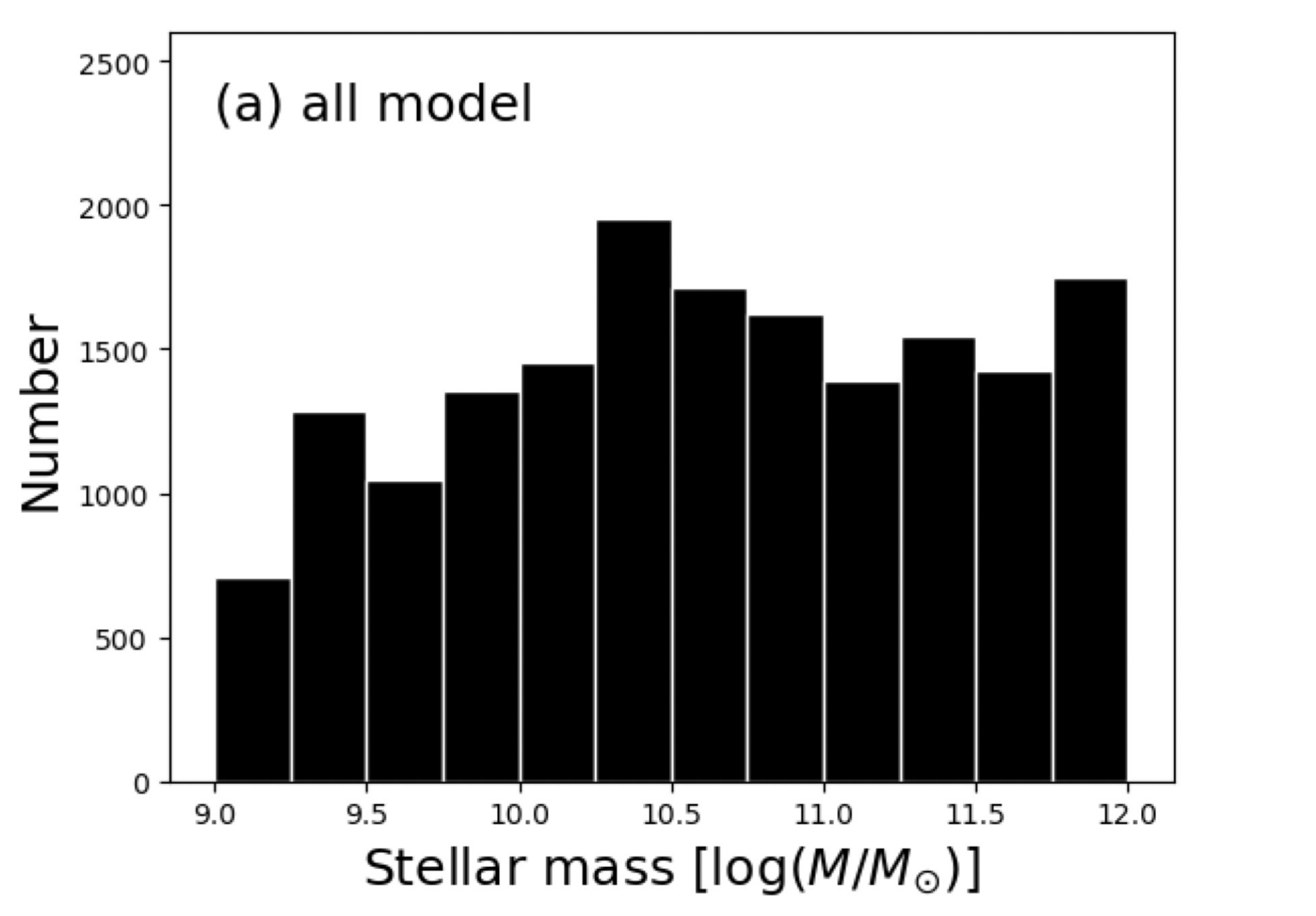} 
    \end{minipage} 
    
     \begin{minipage}[t]{0.33\linewidth}
      \centering 
      \includegraphics[keepaspectratio, width=61mm]{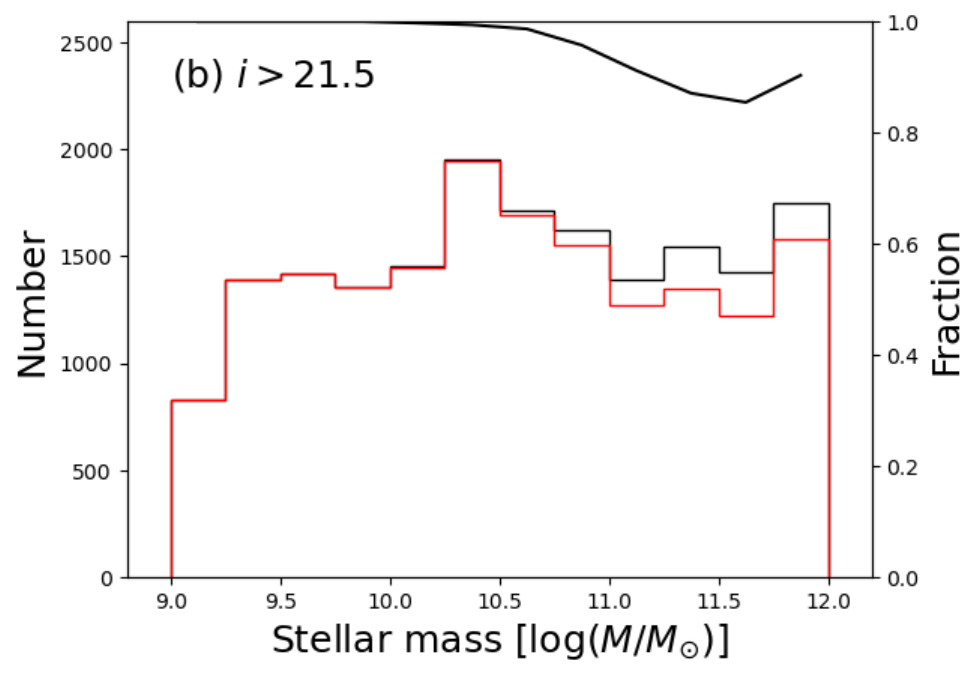} 
    \end{minipage} 
    
    \begin{minipage}[t]{0.33\linewidth}
      \centering
      \includegraphics[keepaspectratio, width=61mm]{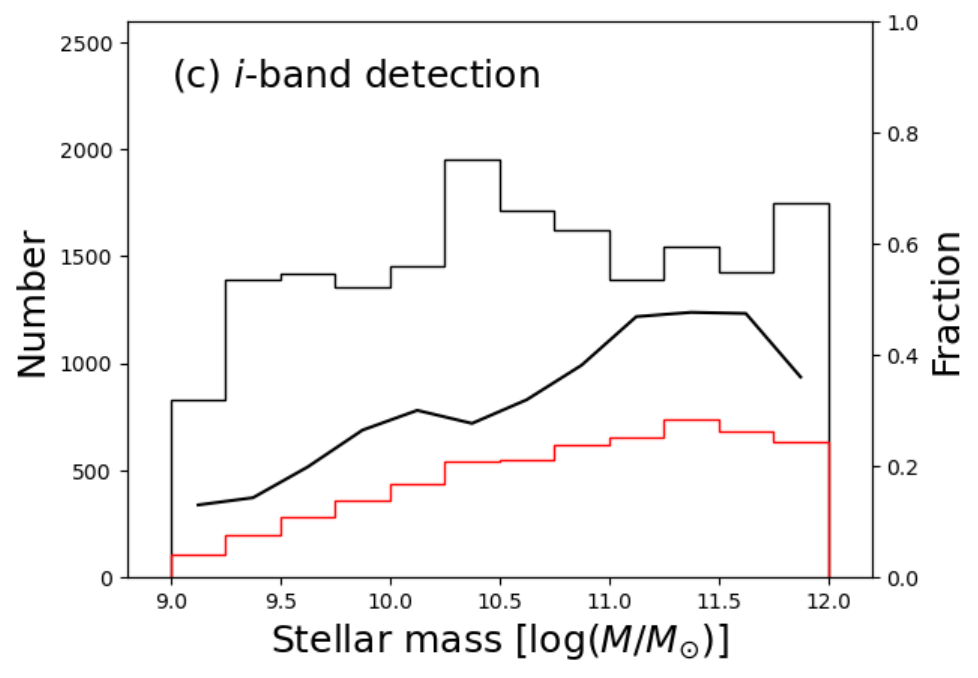}
      \end{minipage} \\

       \begin{minipage}[t]{0.33\linewidth}
      \centering
      \includegraphics[keepaspectratio, width=61mm] {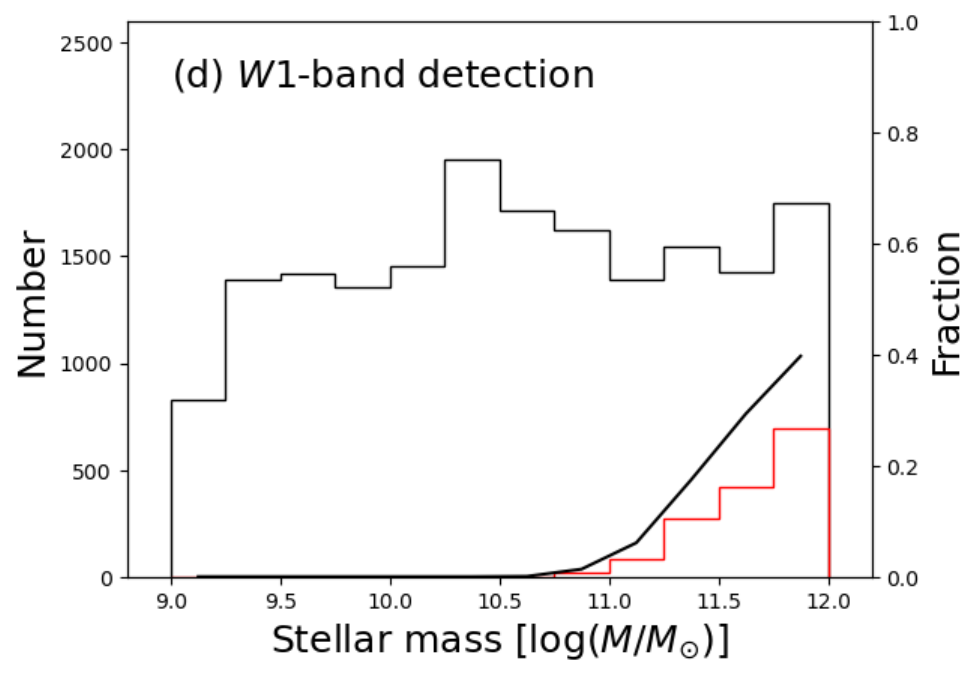}
      
      \end{minipage}

    \begin{minipage}[t]{0.33\linewidth}
      \centering
      \includegraphics[keepaspectratio, width=61mm]{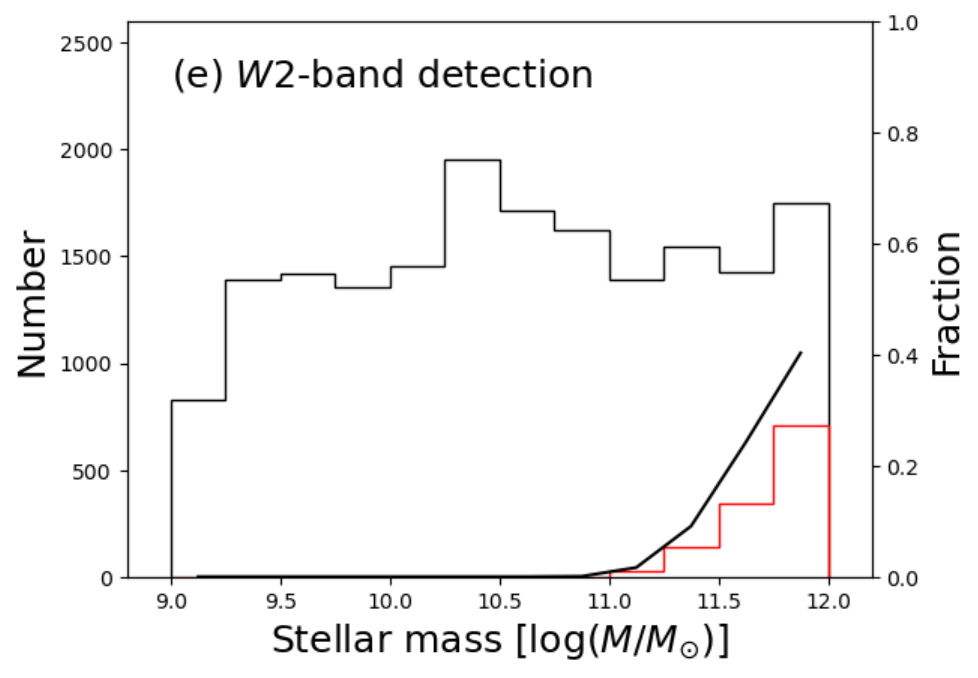}
      \end{minipage}  
      
      \begin{minipage}[t]{0.33\linewidth}
      \centering
      \includegraphics[keepaspectratio, width=61mm] {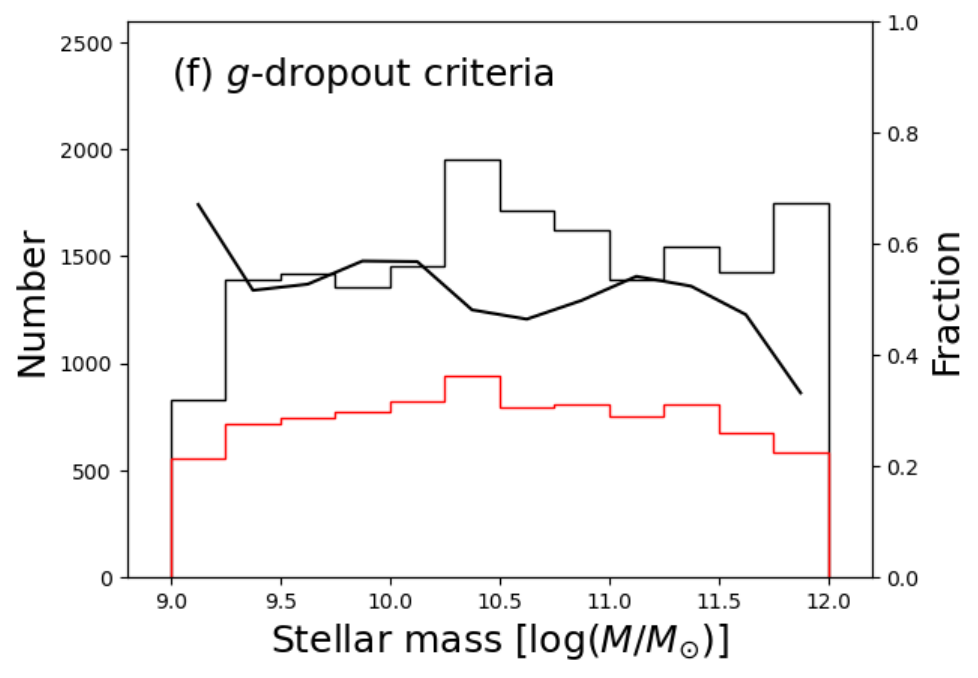}


    \end{minipage} 
  \end{tabular}
  \caption{(a) Number of all models as a function of the stellar mass of galaxies. Panels (b)--(f) show the number of input models (black histogram), models satisfying the criterion (red histogram), and the fraction of galaxies satisfying the criterion (black solid line), as a function of the stellar mass: (b) bright magnitude limit, (c) $i$ band magnitude limit, (d) 3.4 $\mathrm{\mu m}$ magnitude limit, (e) 4.6 $\mathrm{\mu m}$ magnitude limit, and (f) $g$-dropout criteria. }
  \label{fig:seceff}
\end{figure*}

The derived physical parameters of HzRGs presented in Section \ref{subsec:SEDp} are interesting to study the stellar population and star-formation history of HzRGs. However, we may discover only HzRGs in a specific range of the stellar mass ($M_\star$) owing to various selection effects. Therefore, before discussing the outcomes of the SED fit, we assessed such possible selection effects by using X-CIGALE. Specifically, we created various spectral models of galaxies at $z=4$ and applied various selection criteria to estimate the number of galaxies that were selected as a function of the stellar mass. The spectral models were created by adopting the delayed star-formation model with $\tau=10 - 1500 \ \mathrm{Myr}$, $\mathrm{age}= 100 - 1500 \ \mathrm{Myr}$, and $E(B-V)=0.01-2.0$. Consequently, we obtained 17856 models with the stellar mass of 9.0 $<$ log $M_\star/M_\odot$ $<$ 12.0 with an average of log $M_\star/M_\odot = 10.6$ and standard deviation of 0.8. The distribution of the stellar mass of these models is shown in Figure \ref{fig:seceff}a. Subsequently, the following criteria were applied to the spectral models to examine the completeness related to each criterion;
\begin{itemize}
  \item bright magnitude limit ($i = 21.5$) to remove contaminations (see Section \ref{subsubsec:Contami}),
  \item 5$\sigma$ $i$-band limiting magnitude,
  \item 5$\sigma$ infrared limiting magnitude ($W1 < 20.6$, $W2 < 20.1$),
  \item the $g$-dropout criteria (see Section \ref{subsubsec:LBG}).
 \end{itemize}

Figure \ref{fig:seceff} shows the estimated completeness related to each criterion, as a function of the stellar mass. Based on the condition of $i>21.5$ that excludes optically-bright objects, some massive galaxies with $10^{11-12} M_\odot$ were removed; however, its fraction was not high ($\lesssim 0.1$; see Figure \ref{fig:seceff}b). The optical magnitude affected the selection of galaxies in all mass range (9 $<$ log $M_\star/M_\odot$ $<$ 12); as shown in Figure \ref{fig:seceff}c, the completeness was moderate ($\sim$0.4--0.5) in the massive range (log $M_\star/M_\odot$ $\gtrsim$11) but it was lower ($\sim$0.2--0.4) in the less-massive range (log $M_\star/M_\odot$ $\lesssim$11). The selection completeness related to the $g$-dropout criteria did not strongly depend on the stellar mass ($\sim$0.4--0.6; Figure \ref{fig:seceff}f). However, the mid-infrared magnitude limits strongly affected the selection bias, as shown in Figures \ref{fig:seceff}d and \ref{fig:seceff}e. As evident, galaxies with $M_\star < 10^{11} M_\odot$ were not selected. Therefore, it must be considered that our HzRG sample is affected by a strong selection effect in the sense we select only HzRGs with $M_\star > 10^{11} M_\odot$. 

\subsection{Stellar mass and dust extinction} \label{subsec:SM}

\begin{figure}
\begin{center}
 \includegraphics[width=8.0cm]{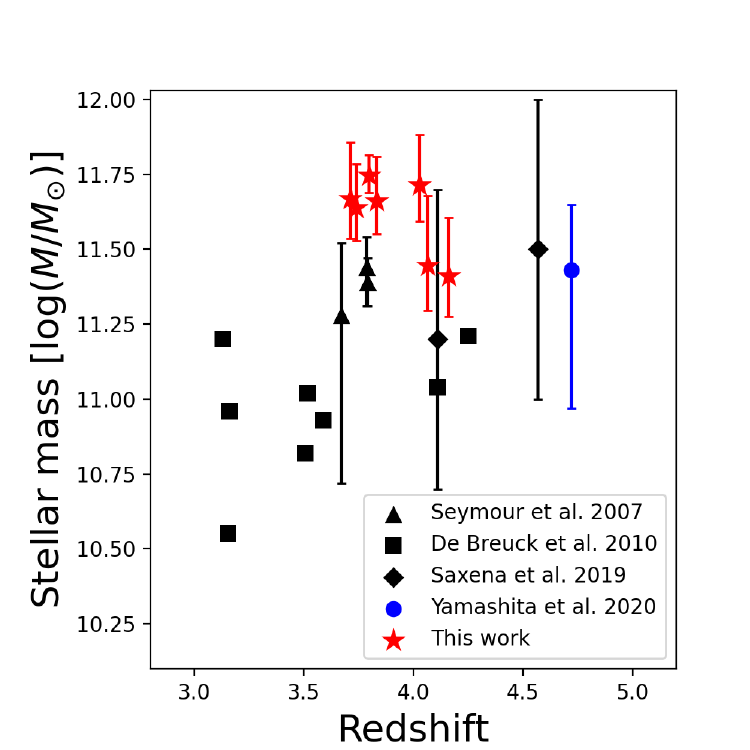}
 \end{center}
 \caption{Stellar mass and redshift for the final HzRG candidate sample (red stars). The blue filled circle denotes the HzRG selected by the $r$-dropout Lyman break technique \citep{2020AJ....160...60Y}. The black plots denote the stellar mass of HzRGs (triangles; \citealt{2010ApJ...725...36D}, squares; \citealt{2007ApJS..171..353S}, diamonds; \citealt{2019MNRAS.489.5053S}) selected by the USS method. }
\label{fig:mass}
\end{figure}

The SED fit with X-CIGALE indicated that the stellar mass of our HzRG sample ranged from $2.6 \times 10^{11} \ M_{\odot}$ to $5.6 \times 10^{11} \  M_{\odot}$, with an average mass of $4.2 \times 10^{11} \ M_{\odot}$. In Figure \ref{fig:mass}, the stellar mass of a HzRG at $z=4.72$ selected through the $r$-dropout Lyman-break technique (($2.6 \pm 1.7) \times 10^{11} M_\odot$; \citealt{2020AJ....160...60Y}) is shown; it was similarly massive to our g-dropout HzRG sample. Figure \ref{fig:mass} also shows the stellar mass of 13 USS-selected HzRGs at similar redshift ($3.0 < z < 5.0$; \citealt{2007ApJS..171..353S, 2010ApJ...725...36D, 2019MNRAS.489.5053S}), which ranged from $0.4 \times 10^{11} \ M_{\odot}$ to $3.2 \times 10^{11} \ M_{\odot}$ with an average mass of $1.5 \times 10^{11} \ M_{\odot}$.  \cite{2007ApJS..171..353S} and \cite{2010ApJ...725...36D} estimated the stellar mass of HzRGs using the PEGASE spectrophotometric model \citep{1997A&A...326..950F} and \cite{2019MNRAS.489.5053S}  employed the python package smpy to estimate the stellar mass. There were some ($>$10) HzRGs with an upper limit on the stellar mass in the redshift range of $3.0 < z < 5.0$ \citep{2007ApJS..171..353S, 2010ApJ...725...36D, 2019MNRAS.489.5053S} owing to insufficient observational limits. These were not considered in the discussion here.
As shown in Figure \ref{fig:mass}, the stellar mass of HzRGs selected via the Lyman-break technique appeared to be somewhat more massive than that of USS-selected HzRGs. To check whether the stellar mass is statistically different between HzRG samples selected by the Lyman-break technique and USS criterion, we performed the Kolmogorov-Smirnov (KS) test. As a result, we obtained a $p$-value of $1.2 \times 10^{-3}$, suggesting that the difference in the mass distributions of the two HzRG samples were statistically significant. This difference was plausibly caused by the selection effect in this study described in Section \ref{subsec:Se}; that is, the magnitude limit in near-infrared and mid-infrared facilitated the sampling of HzRGs with $M_\star > 10^{11} M_\odot$.
Future deeper infrared surveys are required for fair investigations of the stellar-mass distribution of HzRGs selected through the Lyman-break technique.

Here we compare the inferred stellar mass of theHzRG candidates with the stellar mass function of galaxies (not only HzRGs) at $z \sim 4$ (\citealt{2017A&A...605A..70D}). \citet{2017A&A...605A..70D} obtained the mass function for galaxies in the redshift range of $3.5 < z <  4.5$ including both star-forming and passive galaxies in the COSMOS field; the characteristic mass was $\mathrm{log} \ (M_{*} / M_{\odot}) = 11.10 \pm 0.21$. The stellar mass of our HzRG candidate sample ($(4.2 \pm 1.7) \times 10^{11} \ M_{\odot}$ on average) was $\sim$3 times larger than this characteristic mass. Thus, our HzRG sample corresponds almost to the massive end of the mass distribution of galaxies at $z \sim 4$. 

The average of $E(B-V)$ of our HzRG candidate sample was $0.5 \pm 0.2$. 
\citet{2018MNRAS.476.3218C} formulated an empirical relation between the dust extinction and the stellar mass for star-forming galaxies at $z \sim 3.5$. By applying this empirical relation to our HzRG candidate sample, the mean stellar mass of $4.2 \times 10^{11} \ M_{\odot}$ predicted the dust extinction of $E(B-V) = 0.37^{+2.3}_{-2.0}$ considering the \cite{2000ApJ...533..682C} estimation of $R_{V}= 4.05\pm0.80$. This almost corresponded to the $E(B-V)$ inferred from the X-CIGALE analysis (Table \ref{table:data_bayesfit}). Thus, the nature of the dust extinction of our HzRG sample was almost consistent to star-forming galaxies in the similar redshift range. However, the absolute values of the star formation rate of the HzRG estimated in this study can be highly uncertain (see Section \ref{subsec:SEDA}) and thus further discussion on this topic is not straightforward.

\subsection{Rest-frame UVJ color}  \label{subsec:UVJ}

\begin{figure}
 \begin{center}
 \includegraphics[width=8.0cm]{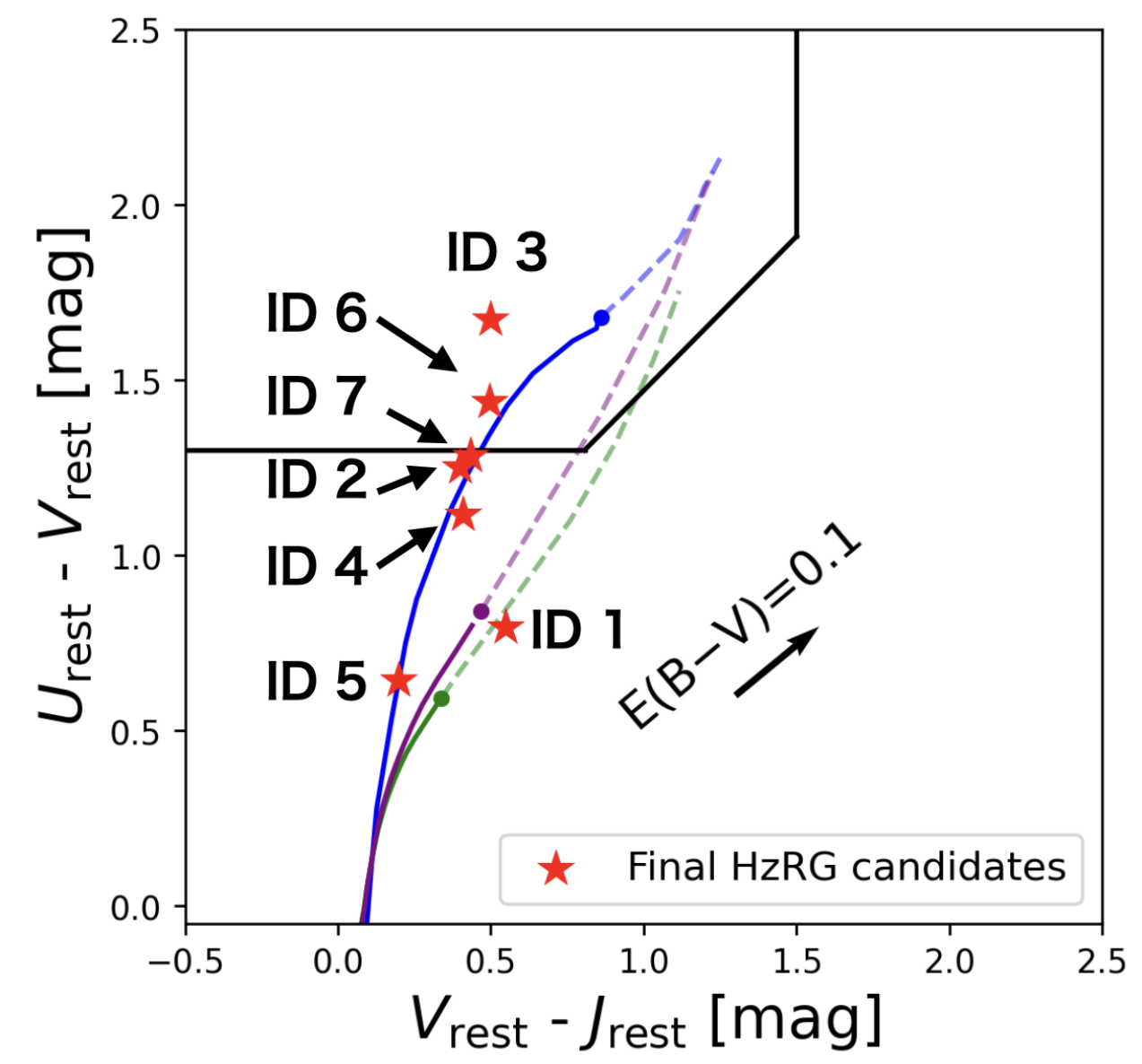}
 \end{center}
 \caption{Rest-frame UVJ diagram. 
 The red stars denote the colors of the 7 HzRG candidates, estimated using the best-fit X-CIGALE models. The black line is the boundary for selecting passive galaxies reported by \citet{2013ApJ...777...18M}. Galaxy evolutionary tracks of BC03 are also shown with $\tau=$ 0.1, 0.5, and 1.0 Gyr (blue, purple, and green, respectively) by adopting the delayed star-formation history with the Solar metallicity and no dust absorption. The model tracks are shown using solid lines up to the age of 1.5 Gyr, which corresponds to the cosmic age at $z \sim 4$. Furtehr, the tracks up to 7 Gyr are shown with dashed lines for references. In addition, the reddening vector of $E(B-V)=0.1$ is shown with a black arrow.
 }
\label{fig:uvj}
\end{figure}

In order to investigate the stellar population of HzRGs, we employed the rest-frame UVJ color diagram (e.g., \citealt{2005ApJ...624L..81L, 2007ApJ...655...51W, 2009ApJ...691.1879W, 2013ApJ...777...18M}). This diagram is widely used to select quenched galaxies effectively, by focusing on the 4000\AA-break in the SED. The UVJ diagram is useful also to investigate how rapidly the past quenching process occurred (e.g., \citealt{2019ApJ...874...17B, 2022A&A...666A.141M}).
The rest-frame UVJ magnitudes of our HzRGs cannot be measured directly from the observed photometric data. This is because the rest-frame $J$-band is beyond the WISE W2 band. Therefore we estimated the UVJ magnitudes of our HzRG candidates from the best-fit galaxy spectral model in the SED fitting process. The rest-frame UVJ colors of our HzRG candidates, obtained in this manner, are presented in Figure \ref{fig:uvj}. The criteria to select quenched galaxies shown by \citet{2013ApJ...777...18M} are as follows; $U-V > 1.3$, $V-J < 1.5$, and $U-V > (U-J) \times 0.88 + 0.59$. Among our HzRG candidates, ID 3 and 6 satisfied the criteria of the quenched galaxy while the remaining 5 HzRG candidates did not. Notably, the UVJ color of ID 2, 4, and 7 was close to the boundary, suggesting that their star-forming activity was not completely quenched but close to the quenched status. 

\begin{figure}
 \begin{center}
 \includegraphics[width=8.5cm]{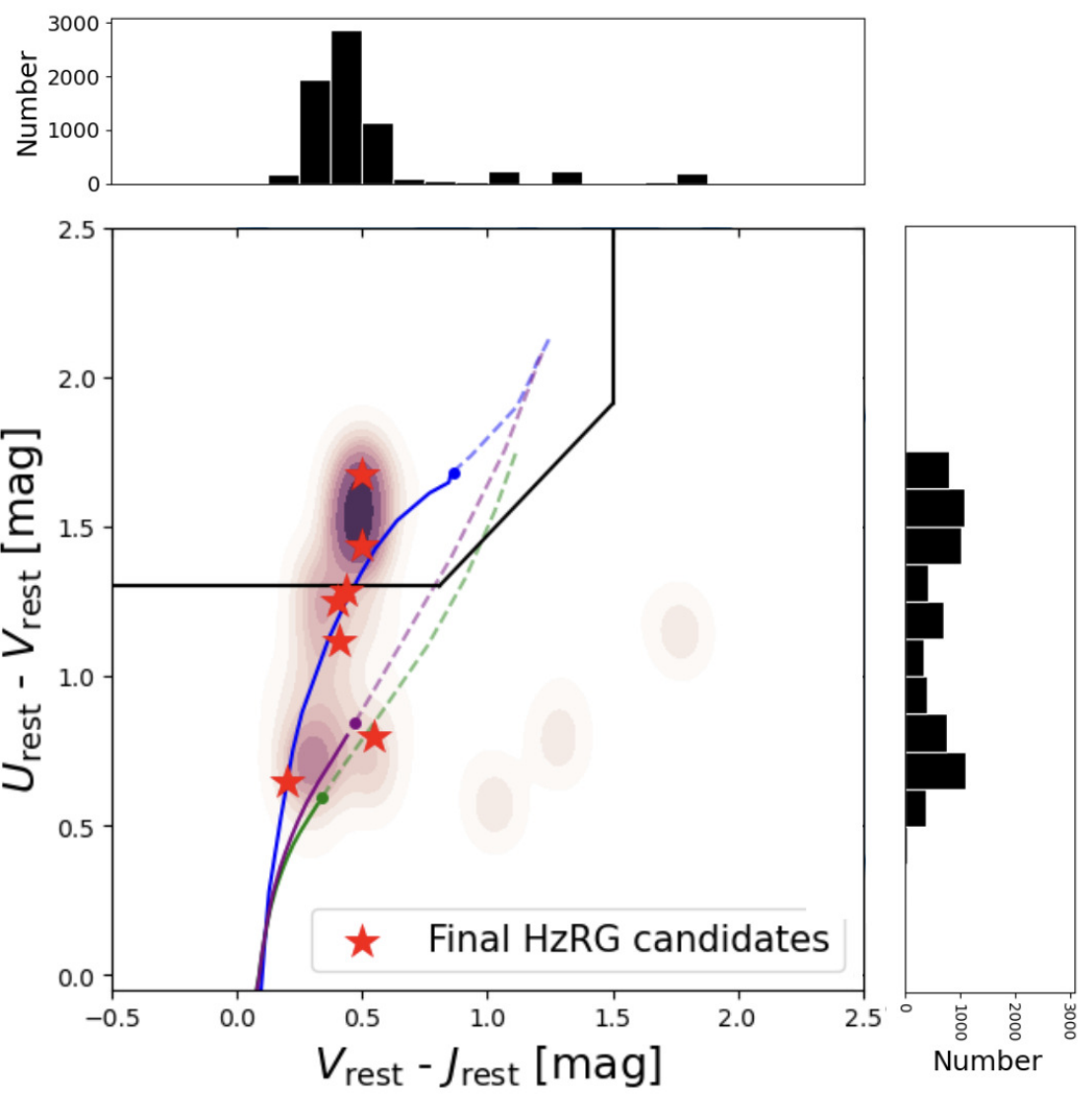}
 \end{center}
 \caption{Same as Figure \ref{fig:uvj} but with the distribution of the UVJ colors of the 7,000 simulated data (see the main text). The histograms of the projected colors are shown on the upper and right sides.
 }
\label{fig:uvj2}
\end{figure}

Till date, various physical mechanisms have been proposed for the fast and slow quenching of the star formation in galaxies. One fast-quenching process is the supernova feedback, which is a gas-removal process owing to the starburst-driven superwind with a timescale of $\sim$0.1 Gyr (e.g., \citealt{2003MNRAS.344.1131D, 2009ApJ...695..292C}). However, this quench mechanism is considered to work mostly in low-mass galaxies (e.g., \citealt{2002MNRAS.337.1299C, 2013MNRAS.428.2741W}). Thus, this process is probably not adequate to explain the quench in our HzRGs whose stellar mass is very massive ($> 10^{11} M_\odot$; Table \ref{table:data_bayesfit}).  A more plausible process of the quenching in massive galaxies is the quasar-mode AGN feedback, which rapidly ($<$0.1 Gyr) suppresses the star formation owing to powerful radiative or kinematic energy of quasars (e.g., \citealt{2005MNRAS.361..776S, 2010MNRAS.402.1516K, 2021MNRAS.507.3985S}). Another possible mechanism that realizes the quenching in galaxies more slowly is the radio-mode AGN feedback, whose typical timescale exceeds 1 Gyr (e.g., \citealt{2005MNRAS.362...25B,2006MNRAS.365...11C}). This process is caused by relativistic jets seen in radio-loud AGNs, thus one plausible mechanism that quenches the star formation in radio galaxies.  To provide a constraint on the physics governing the star-formation history of HzRGs through the quenching timescale, we investigated the galaxy evolutionary track in the UVJ diagram by adopting the BC03 model (Section \ref{subsec:SEDp}). Here we adopted the delayed star-formation models as adopted in Sections \ref{sec:ss} and \ref{sec:results}, with $\tau = 0.1, 0.5,$ and 1.0 Gyr. We adopted the Solar metal abundance as it gives no significant effects on the color track in the range of $0.1 \lesssim Z / Z_\odot \lesssim 1.0$. The results shown in Figure \ref{fig:uvj} indicate that the rest-frame UVJ color of HzRGs were mostly explained by the models with $\tau = 0.1$ Gyr, which is consistent to fast quenching models (e.g., \citealt{2019ApJ...874...17B}). However, without considering the uncertainty in the rest-frame UVJ colors inferred from the best-fit SED models, it is unclear how this result is conclusive. Therefore, we investigated the uncertainty of the UVJ color of the 7 HzRG candidates, by performing the SED fitting for 1,000 sets of photometric data simulating each of the 7 HzRG candidates with varying magnitudes within the photometric error (i.e., 7,000 simulated datasets were created in total). Figure \ref{fig:uvj2} shows the distribution of the UVJ colors of the simulated 7,000 models. As evidently shown, fast quenching was preferred rather than slow quenching, even when considering the uncertainty in the UVJ colors of HzRGs. 

\begin{figure}
 \begin{center}
 \includegraphics[width=8.0cm]{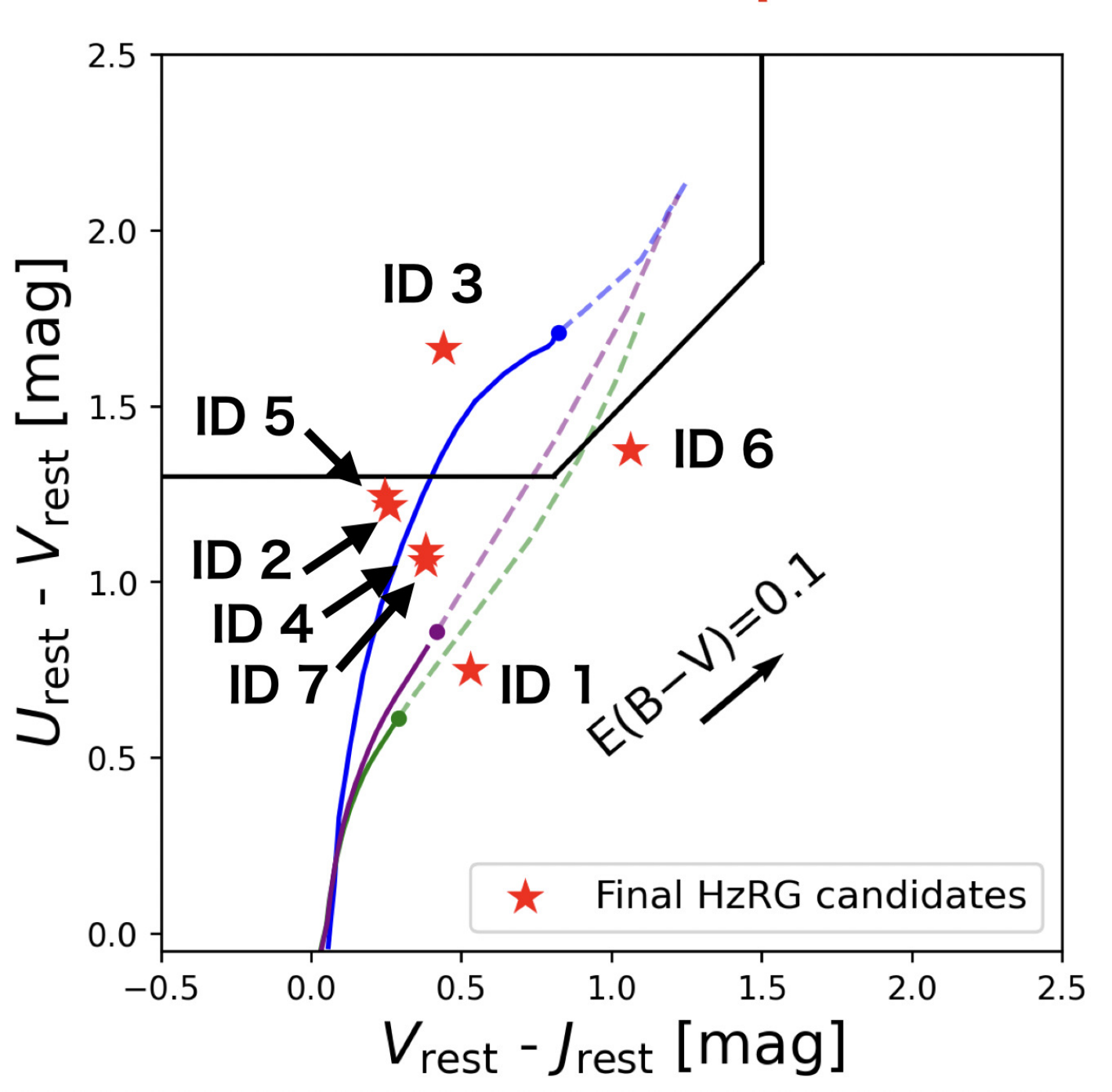}
 \end{center}
 \caption{Same as Figure \ref{fig:uvj} but based on the sfh2exp models instead of the sfhdelayed models (see the main text).}
\label{fig:uvjexp}
\end{figure}

\begin{figure}
 \begin{center}
 \includegraphics[width=8.5cm]{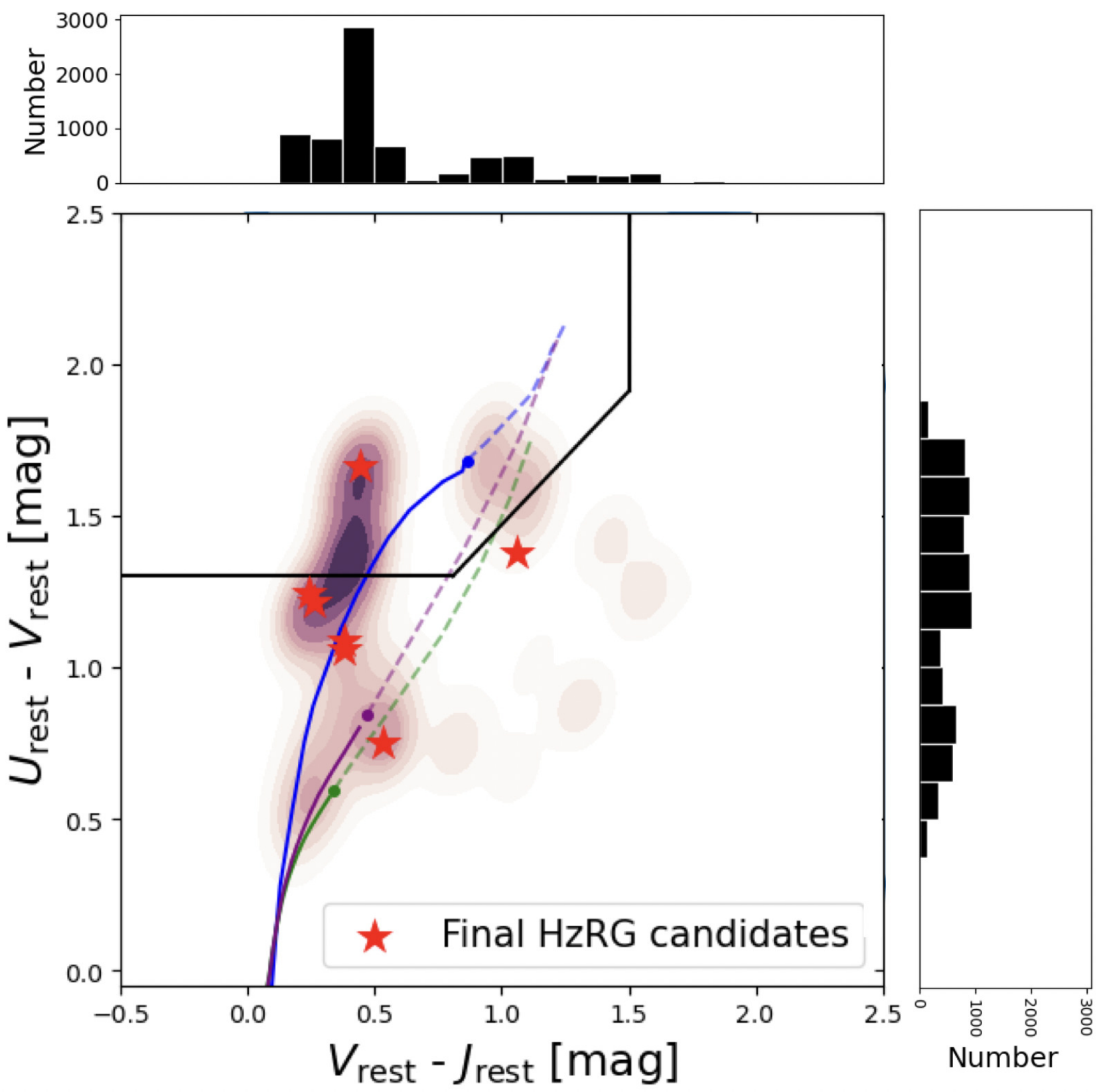}
 \end{center}
 \caption{Same as Figure \ref{fig:uvj2} but based on the sfh2exp models instead of the sfhdelayed models (see the main text).}
\label{fig:uvjexp2}
\end{figure}

However, because of insufficient photometric data for the accurate SED analysis, we adopted some assumptions in the SED fitting procedure (see Section \ref{subsec:SEDA}). To check the robustness of our discussion on the UVJ diagram, we investigated the SED fit adopting the double-exponential SFH model (sfh2exp; see \citealt{2019A&A...622A.103B}) instead of the delayed SFH model (sfhdelayed). As a result, the best-fit sfh2exp models adequately described the observed photometric data of the 7 HzRG candidates with the average $\chi^2 = 1.27$, which is similar to the best-fit sfhdelayed models ($\chi^2 = 1.10$; see Table \ref{table:data_photoz}). Figure \ref{fig:uvjexp} shows the UVJ diagram by adopting the sfh2exp models for both the 7 HzRG candidates and the galaxy evolutionary tracks, while Figure \ref{fig:uvjexp2} shows the uncertainty in the UVJ colors estimated in the same method as in Figure \ref{fig:uvj2}. There is a significant change in ID 6, but the results are in line with the fast quenching track for objects except for ID 1 and ID 6. The similar results in the two different SFH models would reinforce the results.

\subsection{Radio properties}  \label{subsec:radion}

Here we discuss the radio property of the HzRG candidates, by focusing specifically on the radio spectral index and radio luminosity.
The radio spectral index was obtained by combining the FIRST 1.4 GHz data and the 3 GHz data of the Very Large Array Sky Survey (VLASS; \citealt{2020PASP..132c5001L}).
VLASS is a radio survey observed with VLA, conducted at 3 GHz. The angular resolution of the VLASS data is $\sim$2.5 arcsec and the positional accuracy is better than 1.0 arcsec. The version 3 of the Quick Look epoch 1 catalog was used, with Duplicate\_flag $<$ 2, Quality\_flag $== (0 | 4)$, and $S_{\mathrm{code}} \neq E $ as clean sample conditions\footnote[2]{Duplicate\_flag is a flag to denote the duplicate status of the radio source. Quality\_flag is a flag to deal with spurious detections and duplicate objects due to the overlap between tile edges.  $S_{\mathrm{code}} \neq E $ is a flag that excludes sources with empty data.}, which were adopted for making a clean sample of VLASS radio sources. The matching radius between HSC-SSP and VLASS sources was 1 arcsec. As a result, 4 objects (ID 1, 3, 5, and 6) among the 7 final HzRG candidates matched with VLASS sources. Though the VLASS counterpart of ID 7 was not found in the 1 arcsec matching, its VLASS image clearly shows double radio sources with fluxes of $9.84\pm0.34$ mJy (at 1.4 arcsec from the HSC position) and $4.38 \pm 0.31$ mJy (at 2.9 arcsec separation). Thus the sum of these two VLASS soruces was treated as the VLASS flux of ID 7, that is $14.22 \pm 0.46$. The remaining two objects (ID 2 and 4) were not detected in the VLASS image. 

The radio spectral index derived for the observed frequency range between 1.4 GHz and 3.0 GHz ($\alpha_{3000}^{1400}$) was obtained from the FIRST and VLASS fluxes, and is summarized in Table \ref{table:radiopp}. The radio spectral index of ID 2 and 4 was estimated by adopting the VLASS 3 sigma upper limit of 0.36 mJy. The average of $\alpha_{3000}^{1400}$ for the VLASS-detected 5 objects is $\alpha^{1400}_{3000} = -1.09$, while the median value also by taking the two VLASS-undetected sources into account (i.e., for the 7 HzRG candidates) is $-1.32$.
 Among the 7 HzRGs,4 objects (ID 1, 2, 4, and 5)  show the USS characteristic with $\alpha<-1.3$ (e.g., \citealt{2000A&AS..143..303D, 2018MNRAS.480.2733S}) while 3 objects (ID 3, 6 and 7) do not satisfy the USS criterion. This suggests that HzRG searches based on the USS method can miss a fraction of HzRGs, and the Lyman-break method is a powerful complementary way to assess the HzRG population.

\begin{table*}[t]
\begin{center}
\caption{Radio properties of the 7 HzRG candidates}
\label{table:radiopp}
\begin{tabular}{cccccc} \hline

   ID & FIRST flux (mJy)& VLASS flux (mJy) & $\alpha^{1400}_{3000}$ & $L_{\mathrm{1.4 GHz}}$ ($10^{27}\ \mathrm{W\ Hz^{-1}}$) & $L_{\mathrm{3 GHz}}$ ($10^{27}\ \mathrm{W\ Hz^{-1}}$) \\ \hline \hline

1& $16.60\pm0.14$&$6.06\pm0.31$&$-1.32$&$4.47\pm0.04$&$1.63\pm0.08$ \\
2& $1.08\pm0.15$  &$< 0.36^{1}$                    &$<-1.44$&$0.11\pm0.01$&--- \\
3& $12.71\pm0.16$&$8.61\pm0.35$&$-0.51$  &$0.78\pm0.01$&$0.53\pm0.02$ \\
4& $1.22\pm0.15$  &$< 0.36^{1}$                     &$<-1.60$&$0.12\pm0.02$&--- \\
5& $21.39\pm0.14$&$7.20\pm0.31$ &$-1.43$&$5.35\pm0.04$&$1.80\pm0.08$ \\
6& $23.82\pm0.10$&$10.21\pm0.30$&$-1.08$&$3.67\pm0.02$&$1.61\pm0.05$ \\
7& $33.16\pm0.15$&$14.22\pm0.46^{2}$&$-1.11$&$5.47\pm0.02$&$2.34\pm0.08$ \\

    \hline
\end{tabular}
\end{center}
\begin{flushleft}
${}^{\mathrm{1}}$ The 3 sigma limiting flux.

${}^{\mathrm{2}}$ This object shows a double-source image, and the flux shown here is the total of the two ($9.84 \pm0.34$ mJy and $4.38 \pm 0.31$ mJy; see the main text).
\end{flushleft}
\end{table*}

%

\begin{figure}
 \begin{center}
 \includegraphics[width=8.5cm]{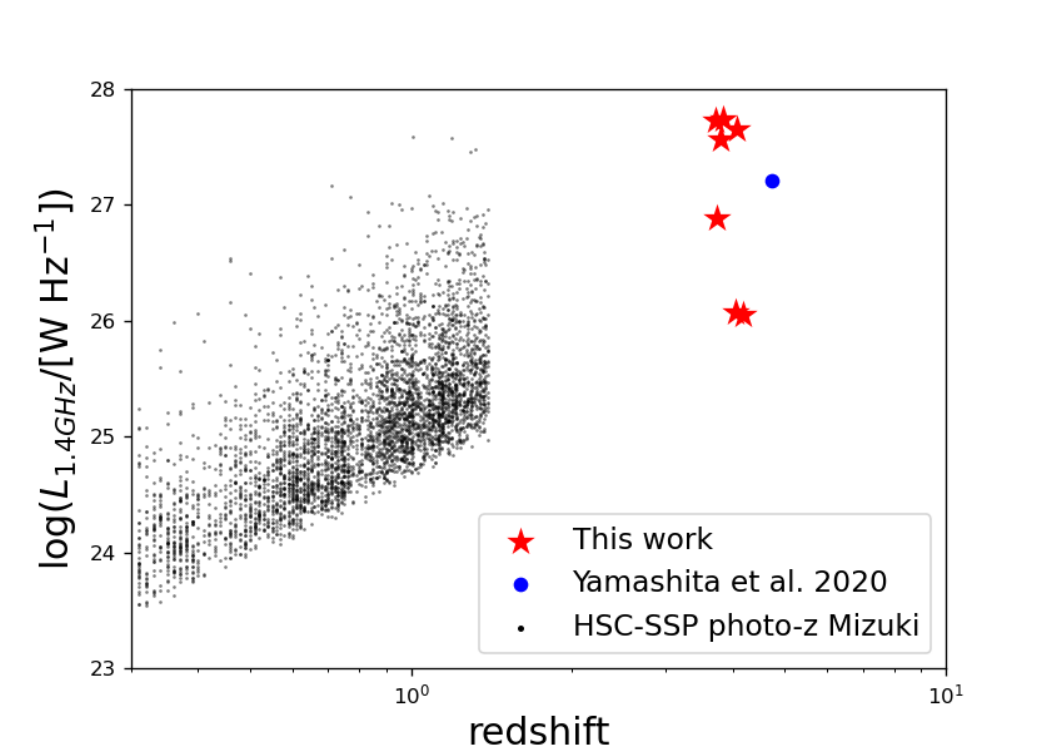}
 \end{center}
 \caption{The rest-frame 1.4GHz radio luminosity as a function of the redshift. The red stars denote the 7 final HzRG candidates. The filled-blue circle denotes the radio galaxy at $z=4.72$ selected by the $r$-dropout Lyman break \citep{2020AJ....160...60Y}. For comparison, the black plots show 4,324 low-$z$ radio galaxies.}
\label{fig:rlumi}
\end{figure}

The luminosity of our HzRG candidates was compared with low-$z$ radio galaxies that were taken from the HSC-SSP Mizuki photo-$z$ catalog (Tanaka 2015; \citealt{2018PASJ...70S...9T}) with FIRST and VLASS detections. The Mizuki photo-$z$ works well particularly in the redshift range of $z \sim 0.3-1.4$, where the HSC filters can detect the Balmer/4000\AA\ break (see, e.g., \citealt{2020ApJ...904..128I}). For limiting the sample to galaxies with good photo-$z$ accuracy, we selected galaxies with photo-$z$ error $\sigma< 0.1$ and reduced $\chi^{2} < 3$, based on \cite{2018ApJ...866..140Y}, \cite{2019ApJS..243...15T}, and \cite{2022ApJ...934...68U}. Through this selection and also requiring the FIRST and VLASS detections, we obtained 4,324 low-$z$ radio galaxy sample at $z = 0.3-1.4$.
The rest-frame radio luminosity of the 4 HzRG candidates having the VLASS counterpart was calculated by adopting the photo-$z$ and the derived radio spectrum index, and given in Table \ref{table:radiopp}. For the other radio galaxies, the radio luminosity was derived by assuming $\alpha=-0.7$ (e.g., \citealt{1992ARA&A..30..575C}). Figure \ref{fig:rlumi} shows the 1.4 GHz Luminosity of HzRGs (including the 7 HzRG candidates and the HzRG at $z=4.72$ reported by \citealt{2020AJ....160...60Y}) and low-$z$ radio galaxies, as a function of the redshift. This figure shows that the radio luminosity of the 7+1 HzRGs corresponds to the luminous end of the luminosity distribution of low-$z$ radio galaxies. Especially, ID 1, 5, 6, and 7 are the most luminous in radio, and thus they are very interesting objects in future studies on the extremely powerful radio jets in the early Universe.

\section{CONCLUSION} \label{sec:conclusion}

In this study, we searched for HzRGs at $z \sim 4$ using $g$-dropout Lyman-break technique, which is free from bias toward steep radio spectral index that has been used in past HzRG searches. Owing to the deep and wide imaging dataset obtained via the HSC-SSP survey, we obtained the following results and implications.
 \begin{itemize}
  \item{We discovered 146 HzRG candidates at $z \sim 4$ which exhibited a significant Lyman-break feature in their SED. Among them, we found 7 HzRG candidates whose photometric redshifts were in the range of $3.3 < z_{\rm ph} < 4.5$; these were derived through the combination of the HSC-SSP, UKIDSS/VIKING, and unWISE survey data.}
  \item{The SED fit with X-CIGALE suggested that the stellar mass of the all of 7 HzRG candidates was very massive; it was distributed in the range of $(2.6-5.6) \times 10^{11} M_\odot$ with the average of $4.2 \times 10^{11} M_\odot$. This is significantly more massive than the characteristic mass of galaxies in the same redshift. Although such a massive nature of HzRGs has been reported for USS-selected HzRGs, our results for HzRGs selected by the near-infrared detected Lyman-break galaxies suggest that the massive nature is a general characteristic of HzRGs that is independent of the selection method. However, as infrared detection contributes to the stellar mass bias, further systematic studies with deeper photometric data are required to clarify the complete shape of the stellar mass distribution function of Lyman-break HzRGs.}
  \item{The rest-frame UVJ diagram indicated that two of 7 HzRG candidates were classified as quiescent galaxies while the remaining HzRGs were also close to be quiescent. The comparison with galaxy evolutionary models indicated that the timescale of the past quenching of HzRGs was fast ($\sim$0.1 Gyr). }
  \item{The radio spectral index of 4 HzRG candidates with a VLASS counterpart is diverse; 4 objects satisfy the USS criterion but the remaining 3 objects do not. This demonstrates the importance of HzRG searches with the Lyman-break method, which can select HzRGs that are missed by USS-based surveys.}
 \end{itemize} 

\vspace{3mm}

\section*{acknowledgments}
The Hyper Suprime-Cam (HSC) collaboration includes the astronomical communities of Japan and Taiwan, and Princeton University.  The HSC instrumentation and software were developed by the National Astronomical Observatory of Japan (NAOJ), the Kavli Institute for the Physics and Mathematics of the Universe (Kavli IPMU), the University of Tokyo, the High Energy Accelerator Research Organization (KEK), the Academia Sinica Institute for Astronomy and Astrophysics in Taiwan (ASIAA), and Princeton University.  Funding was contributed by the FIRST program from the Japanese Cabinet Office, the Ministry of Education, Culture, Sports, Science and Technology (MEXT), the Japan Society for the Promotion of Science (JSPS), Japan Science and Technology Agency  (JST), the Toray Science  Foundation, NAOJ, Kavli IPMU, KEK, ASIAA, and Princeton University. This paper is based in part on data collected at the Subaru Telescope and retrieved from the HSC data archive system, which is operated by Subaru Telescope and Astronomy Data Center (ADC) at NAOJ. Data analysis was in part carried out with the cooperation of Center for Computational Astrophysics (CfCA) at NAOJ.  We are honored and grateful for the opportunity of observing the Universe from Maunakea, which has the cultural, historical and natural significance in Hawaii.
 This paper makes use of software developed for Vera C. Rubin Observatory. We thank the Rubin Observatory for making their code available as free software at http://pipelines.lsst.io/. 
The Pan-STARRS1 Surveys (PS1) and the PS1 public science archive have been made possible through contributions by the Institute for Astronomy, the University of Hawaii, the Pan-STARRS Project Office, the Max Planck Society and its participating institutes, the Max Planck Institute for Astronomy, Heidelberg, and the Max Planck Institute for Extraterrestrial Physics, Garching, The Johns Hopkins University, Durham University, the University of Edinburgh, the Queen's University Belfast, the Harvard-Smithsonian Center for Astrophysics, the Las Cumbres Observatory Global Telescope Network Incorporated, the National Central University of Taiwan, the Space Telescope Science Institute, the National Aeronautics and Space Administration under grant No. NNX08AR22G issued through the Planetary Science Division of the NASA Science Mission Directorate, the National Science Foundation grant No. AST-1238877, the University of Maryland, Eotvos Lorand University (ELTE), the Los Alamos National Laboratory, and the Gordon and Betty Moore Foundation.

This work was supported by JST, the establishment of university fellowships towards the creation of science technology innovation, Grant Number JPMJFS2131.
This work was nancially supported by JSPS KAKENHI grants: Nos. 20H01949 (TN), 21H04490 (HU), 22K14075 (HU), 23H01215 (TN), and 23K22537 (YT). We thank the anonymous referee, whose useful comments improved this paper significantly. 


\bibliography{Ref}{}


\end{document}